\documentclass[fleqn,usenatbib]{mnras}

\usepackage{newtxtext,newtxmath}

\usepackage[T1]{fontenc}

\DeclareRobustCommand{\VAN}[3]{#2}
\let\VANthebibliography\thebibliography
\def\thebibliography{\DeclareRobustCommand{\VAN}[3]{##3}\VANthebibliography}

\usepackage{graphicx}	
\usepackage{amsmath}	
\usepackage{comment}
\usepackage{xspace} 
\usepackage{mathtools}
\usepackage{hyperref}
\usepackage{float}
\usepackage{bm}
\usepackage{pdflscape}
\usepackage{pdfpages}
\usepackage{longtable}
\usepackage{algorithmicx}
\usepackage{algorithm} 
\usepackage{adjustbox}
\usepackage{subfigure}
\usepackage[noend]{algpseudocode}
\DeclareFontFamily{U}{mathx}{}
\DeclareFontShape{U}{mathx}{m}{n}{<-> mathx10}{}
\DeclareSymbolFont{mathx}{U}{mathx}{m}{n}
\DeclareMathAccent{\widecheck}{0}{mathx}{"71}

\usepackage{multirow} \usepackage{threeparttable}
\usepackage{booktabs} \usepackage{stfloats}

\algdef{SE}[DOWHILE]{Do}{doWhile}{\algorithmicdo}[1]{\algorithmicwhile\ #1}%

\DeclareFontFamily{U}{mathx}{\hyphenchar\font45}
\DeclareFontShape{U}{mathx}{m}{n}{<-> mathx10}{}
\DeclareSymbolFont{mathx}{U}{mathx}{m}{n}

\DeclarePairedDelimiterX{\Paren}[1]{(}{)}{#1}
\DeclarePairedDelimiterX{\Brace}[1]{\{}{\}}{#1}
\DeclarePairedDelimiterX{\Brack}[1]{[}{]}{#1}
\DeclarePairedDelimiterX{\Abs}[1]{\rvert}{\lvert}{#1}
\DeclarePairedDelimiterX{\Norm}[1]{\lVert}{\rVert}{#1}
\DeclarePairedDelimiterX{\Avg}[1]{\langle}{\rangle}{#1}
\DeclarePairedDelimiterX{\Round}[1]{\lfloor}{\rceil}{#1}
\DeclarePairedDelimiterX{\Floor}[1]{\lfloor}{\rfloor}{#1}
\DeclarePairedDelimiterX{\Ceil}[1]{\lceil}{\rceil}{#1}
\DeclarePairedDelimiterX{\Inner}[2]{\langle}{\rangle}{#1,#2}

\DeclarePairedDelimiterXPP{\Expect}[1]{\mathbb{E}}(){}{#1}

\usepackage[squaren,Gray]{SIunits}

\usepackage{numprint}

\makeatletter
\def\widebreve{\mathpalette\wide@breve}
\def\wide@breve#1#2{\sbox\z@{$#1#2$}%
     \mathop{\vbox{\m@th\ialign{##\crcr
\kern0.08em\brevefill#1{0.8\wd\z@}\crcr\noalign{\nointerlineskip}%
                    $\hss#1#2\hss$\crcr}}}\limits}
\def\brevefill#1#2{$\m@th\sbox\tw@{$#1($}%
  \hss\resizebox{#2}{\wd\tw@}{\rotatebox[origin=c]{90}{\upshape(}}\hss$}
\makeatletter

\title[Logistic regression to boost exoplanet detection performances]{Logistic regression to boost exoplanet detection performances}

\author[H. Cambazard et al.]{
Hadrien Cambazard$^{1}$\thanks{E-mail: hadrien.cambazard@g-scop.grenoble-inp.fr}, Nicolas Catusse$^{1}$, Antoine Chomez$^{2,3}$, Anne-Marie Lagrange$^{2,3}$
\\
$^{1}$Univ. Grenoble Alpes, CNRS, Grenoble INP, G-SCOP, F-38000 Grenoble, France \\
$^{2}$Laboratoire d'{\'E}tudes Spatiales et d'Instrumentation en Astrophysique, Observatoire de Paris, Univ. PSL, Sorbonne Univ., Univ. Paris Diderot, France\\
$^{3}$Univ. Grenoble Alpes, Institut de Planétologie et d'Astrophysique de Grenoble, France
}

\pubyear{2023}

\begin{document}
\label{firstpage}
\pagerange{\pageref{firstpage}--\pageref{lastpage}}
\maketitle

\begin{abstract}
Direct imaging of exoplanets requires to separate the background noise from the exoplanet signals. Statistical methods have been recently proposed to avoid subtracting any signal of interest as opposed to initial self-subtracting methods based on Angular Differential Imaging (ADI). However, unless conservative thresholds are chosen to claim for a detection, such approaches tend to produce a list of candidates that include many false positives. Choosing high, conservative,  thresholds leads to miss the faintest planets. We extend a statistical framework with a logistic regression to filter the list of candidates. Features with physical/optical  meaning (in two wavelengths) are used, leading to a very fast and pragmatic approach. The overall method requires a simple edge detection (image processing) and clustering algorithm to work with sub-images. To estimate its efficiency, we apply our approach to targets observed with the ESO/SPHERE high contrast imager, that were previously used as tests for blind surveys. Experimental results with injected signals show that either the number of false detections is considerably reduced or faint exoplanets that would otherwise not be detected can be sometimes found. Typically, on the blind tests performed, we are now able to detect around 50\% more of the injected planets with an SNR below 5, and with a very low number of additional candidates.

\end{abstract}

\begin{keywords}
Exoplanets. Techniques: high angular resolution -- techniques: image processing -- methods: data analysis \vspace*{-0.5cm}
\end{keywords}



\section{Introduction}
\label{sec:introduction}

High contrast imaging (HCI) of exoplanets has become an important area in astronomy, as it provides direct information of the exoplanets orbital parameters, as well as measurements of their fluxes, which, in turn, can be used to estimate, through comparisons with evolution models, their masses, effective temperatures, and gravity. Coupled to spectroscopy, HCI also allows for atmosphere characterization. It is recognized as a key technique to study Earth twins in the future. Given today instrumental capabilities, only massive giant planets orbiting further than typically 10 au have been detected, and most of them are young (tens to hundreds of Myr). 

HCI relies on three pillars: adaptive optics (AO), coronagraphy and data processing. Since the first AO imagers in the nineties on 4m class telescopes, two generations of instruments have been developped on 10m class telescopes, with increased AO and coronagraphy capabilities. The first generation led to imaging of a few exoplanets \cite[see e.g.][]{Chauvin04, Marois08, lagrange2009probable}. Second generation instruments using extreme AO and higher performance coronagraphs \citep{beuzit2019sphere,macintosh2014first} have allowed us to detect a few more planets and provide interesting constraints on the massive giant planet demographics beyond 10 au \citep{Nielsen2019, vigan21_shine3}.

While more powerful instruments will allow detecting lighter and closer planets and deriving more accurate and complete information on exoplanets demographics, using better detection algorithms to separate the stellar noise from the exoplanet signal as much as possible can allow for improving exoplanets detection capabilities on today data. A number of techniques have been designed to cancel out the background speckles by combining images taken at consecutive times. The key idea of the ADI technique \cite{osti_887287,Lagrange_2010} is to subtract the temporal median to each frame and realign the residual images to recover the remaining signals of the companions at the same location. This median acts as a reference Point Spread Function (PSF) that models the speckle patterns to be subtracted. 
Various alternatives to this basic approach have been proposed to estimate the reference stellar PSF to be removed such as the LOCI algorithm \cite{marois_correia_veran_currie_2013,Absil2013SearchingFC}, the principal component analysis based algorithms \cite{Soummer2012} or LLSG \cite{GonzalezAbsil2016} which uses a low rank approximation to model the PSF. Yet, the  subtraction of the PSF may lead to a loss of the companion signal because it is fitted in the PSF. This so-called \emph{self-subtraction} is a limitation associated to such approaches. Statistically-based methods then appeared from the need to evaluate the confidence of each detection and have been applied after the speckle cancellation step. For instance, ANDROMEDA \citep{Cantalloube2015} uses a maximum likelihood estimation to detect planetary signals. This estimation is performed under the assumption of a Gaussian distribution of the noise. This approach produces a Signal to Noise Ratio (SNR) map that can be filtered out with a threshold to identify detected sources. A planetary signal shows an expected pattern typically made of a positive oval lobe surrounded by two negative lobes. But the resulting SNR map still contains a number of artifacts whose signals are above the threshold;  three sources of false detections are diagnosed in \cite{Cantalloube2015}: a tertiary lobe in the expected pattern, spider diffraction patterns and high speckle noise in close areas to the star. To find and reject the artifacts, a number of tests are performed by fitting the morphological expected shape of the planetary signal. 
All previously mentioned works rely on unsupervised learning techniques. The lack of a large enough labelled dataset of positive and negative samples is a challenge for supervised learning algorithms. Firstly, the small number of confirmed exo-planets leads to class imbalance with very few guaranteed positive samples. Secondly, images where no candidate planets have been detected can not be used as representative of negative samples since yet undiscovered planets might be present.  

Recently, a random forest and neural network classifier were presented in \cite{Gonzales2018} where the training data is obtained by injecting fake exo-planets and blurring  pre-processed ADI sequence of images. Alternatively, a Generative Adversarial Network is trained in \cite{Yip2020} to get a generative model of the speckle noise pattern and use it to create large negative labeled datasets. Fake planets are then injected in such guaranteed negative samples. A Convolutional Neural Network is then trained and planets are localized using Class Activation Maps (CAM). A post-processing based on half-sibling regression (HSR) is proposed in \cite{gebhard2020physically}. The authors argue that the speckle pattern can be approximately anti-symmetric across the origin in some circumstances which can help to choose the predicators of the HSR framework. Recently, supervised deep learning has been combined with  the statistical model of PACO, leading to a gain of 0.5 mag compared to PACO \citep{flasseur2023}. 

In this paper, we extend the statistical framework of \cite{flasseur2018exoplanet} using a simple logistic regression to filter the list of candidates and simple features expected from a planetary signal and noise. The overall methodology also relies on edge detection and clustering to present a meaningful list of candidate planets to the user. The approach is described in Section 2. We apply it on test cases in Section 3, and analyse the results. We finally use this approach to explore the environment of one iconic system (Section 4).


\section{Approach}
\label{sec:approach}
\subsubsection{Input data}

We use SPHERE/InfraRed Dual-band Imager and Spectrograph (IRDIS) \citep{2008SPIE.7014E..3LD} data obtained in Angular and Spectral Differential Imaging mode (ASDI). 
Both H2 and H3 spectral channels are considered. A pre-processing based on a standard pipeline in astronomy is performed with a number of calibration steps (dark, flat, offset, ...) partly dedicated to the instrument. The data sets contain stellar noise close to the star, and background/detector noise further away. The PACO algorithm for (PAtches COvariances) \cite{flasseur2018exoplanet} is used to model the noise and to produce an SNR map. In this approach, the background variations are considered to be spatially correlated and non-stationary. A statistical model for these fluctuations is proposed in \cite{flasseur2018exoplanet} and the presence of a companion is established using a statistical hypothesis test. More precisely, the background is assumed to follow a multivariate Gaussian distribution locally. The field of view is thus decomposed into small patches that are large enough to contain the core of a point source. A Gaussian distribution is estimated for each patch, to take into account spatial correlation, by using the different images of the patch over time. A hypothesis test is formulated for the pixel at the center of the patch to decide whether a pure background hypothesis can be rejected. The test statistic can be interpreted as a Signal to Noise Ratio (SNR) producing an image or a map that can be inspected (visually or algorithmically) for the presence of companions. Application of PACO to astrophysical sources can be found in \cite{chomez2023a} and in \cite{Chomez2023b}. 

A simple threshold of this SNR map can in principle be used to identify sources. A threshold of 5$\sigma$ has been used so far when analysing the data with PACO. Using lower thresholds would indeed lead to too many candidates (typically hundreds). \cite{Cantalloube2015} considered slightly lower thresholds by adding a number of dedicated tests to reject artifacts. We propose to enrich the threshold with a limited number of features with optical/physical meanings.

\subsection{Principle}
 We use a simple classifier based on logistic regression to filter the list of source candidates. Simple features are computed on the speckle and planet patterns resulting in a very fast and pragmatic algorithm to classify sub-images of the SNR map.
 
 It is similar to \cite{gebhard2020physically} in the sense that scientific knowledge is directly provided to the classifier but it follows a very different implementation. The signal of an exoplanet in the original image is expected to have a specific structure in the SNR map. Typically, an Airy figure in the original image affects the SNR distribution and even if the corresponding structure does not necessarily have the exact characteristics of an Airy Figure, it can still be indicative of the presence of an exoplanet. For sake of simplicity, we shall refer to it as an Airy Figure though this is not technically exact. In addition, a speckle is similar in both wavelengths but appears at different locations in each channel according to a radial shift from the center of the image (the star). A number of such features are computed on small sub-images and a simple classifier is trained to discriminate noise from planetary signals.



\subsection{Methodology}
The overall process starts from the H2 and H3 SNR maps available after processing the original data cube with the PACO algorithm. The process is summarized Figure \ref{fig:process} in five steps:

\begin{figure*}
\centering
\includegraphics[width=16cm]{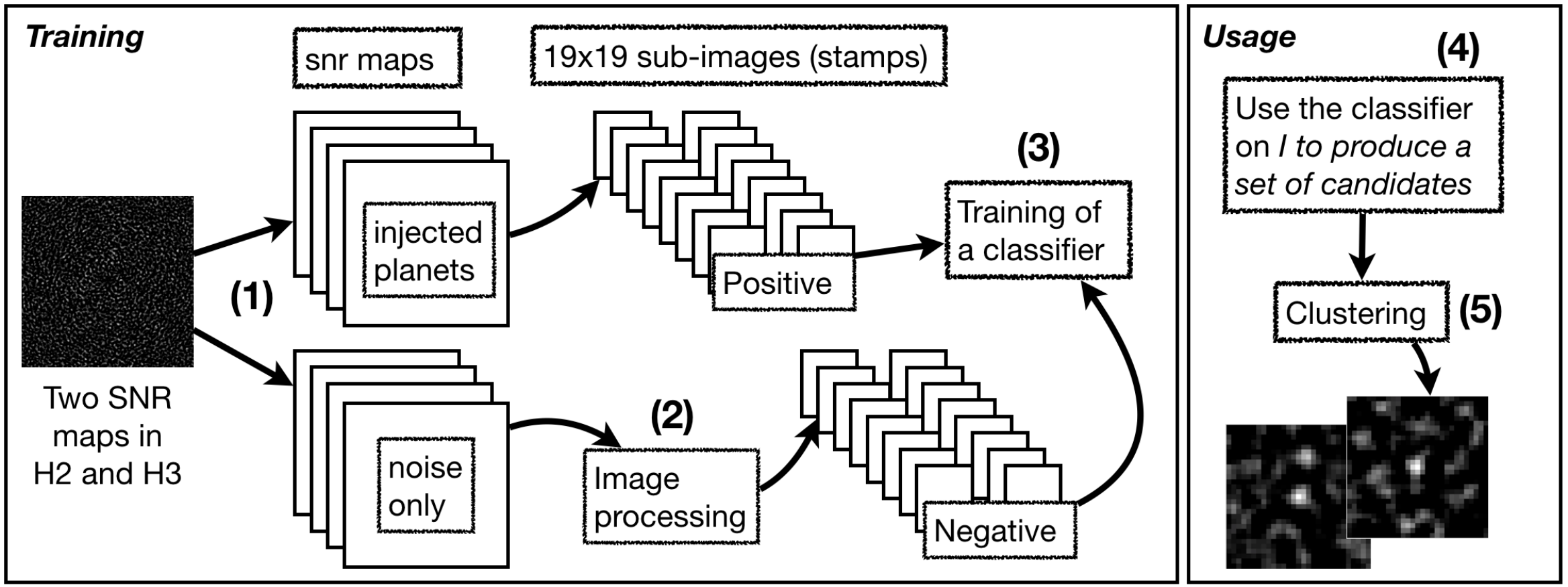}%
  \caption{Overview of the methodology.\label{fig:maps}}
  \label{fig:process}
\end{figure*}

 \begin{enumerate}
 \item (\emph{Training}) Generation of the SNR maps with injected exoplanets or noise. Small sub-images, also referred to as \textbf{stamps}, will be extracted from these guaranteed SNR maps. Typically, a stamp containing an injected exoplanet is considered as a positive stamp (of the positive class). Positive stamps are directly produced from the SNR maps injected with planets by centering the stamp at the location of the injection.
 \item (\emph{Training}) Negative stamps refer to stamps that do not contain a planet and can be centered at any pixel of the noise maps. This would give rise to nearly a million stamps for a single map with a lot of redundancy (overlapping). Since a candidate object can only be detected at a peak of SNR value we use image processing techniques to center the stamps on SNR peaks (see Section \ref{stampgene}). This step produces a collection of negative stamps extracted from the SNR maps guaranteed with noise only.
 \item (\emph{Training}) A classifier is trained from the collection of stamps whose class is known. Note that the classifier is thus dedicated to a specific image that was used to produce the SNR maps with planets/noise.
 \item (\emph{Usage}) For each pixel $u$ of SNR intensity greater than a threshold value (in practice set to 2), a stamp centered on $u$ is submitted to the classifier. This step produces a collection of \textbf{candidates}.
 \item (\emph{Usage}) The stamps classified as possible candidates are then clustered (since they can considerably overlap) and the set of clusters is presented to the user. 
 \end{enumerate}
 
\subsubsection{Step (1):Generation of the training data-set: planet's injections and noise map}
\label{sec:injection}
A data cube is used to generate a number of SNR maps (see Figure \ref{fig:maps}) that contains either only noise or signals of previously injected fake giant planets of various  masses. The objective is to provide guaranteed positive and negative samples for the machine learning algorithm.

\medskip


{\bf Noise maps.} There are two ways to create a "noise map only" from the data set. The first one is to invert the direction of rotation. Because we use ADI-based techniques taking advantage of the rotation, inverting the direction destroys any astrophysical signal, leaving only noise. Another way to create noise only maps is to perform a temporal shuffle of the frames.

\medskip
{\bf Injection of fake planets.} 
Fake planets (FPs) with given masses (expressed in Jupiter masses $\text{M}_{\text{jup}})$ are randomly injected in the reduced and centered data cubes provided by the SPHERE data center \citep{delorme17sphere}, \citep{chomez2023a} taking into account the inverted direction of rotation (see above). The injected masses range from 1 to 5 $\text{M}_{\text{jup}}$. A minimum distance between the injections is enforced to avoid any stacked signal.
To convert masses into contrasts in the H2 and H3 bands, we use COND \citep{COND_model} evolution models, and assume that the planet is coeval with its parent star.
Those data cubes are then processed by PACO using ADI and ASDI.

\medskip
Each  SNR map generated from an original image $\mathcal{I}$ (made of two SNR maps in H2 and H3) gives a number of sub-images (\emph{stamp}) that are known to contain or not an object (an injected planet/companion). The size of the stamps, 19x19 pixels, is large enough to give the context needed to detect an astrophysical point source (field companion or exoplanet). 

\subsubsection{Step (2): Negative stamps from the noise maps}
\label{stampgene}
\begin{figure*}
  \centering
\begin{subfigure}
  \centering
  \includegraphics[width=.4\linewidth]{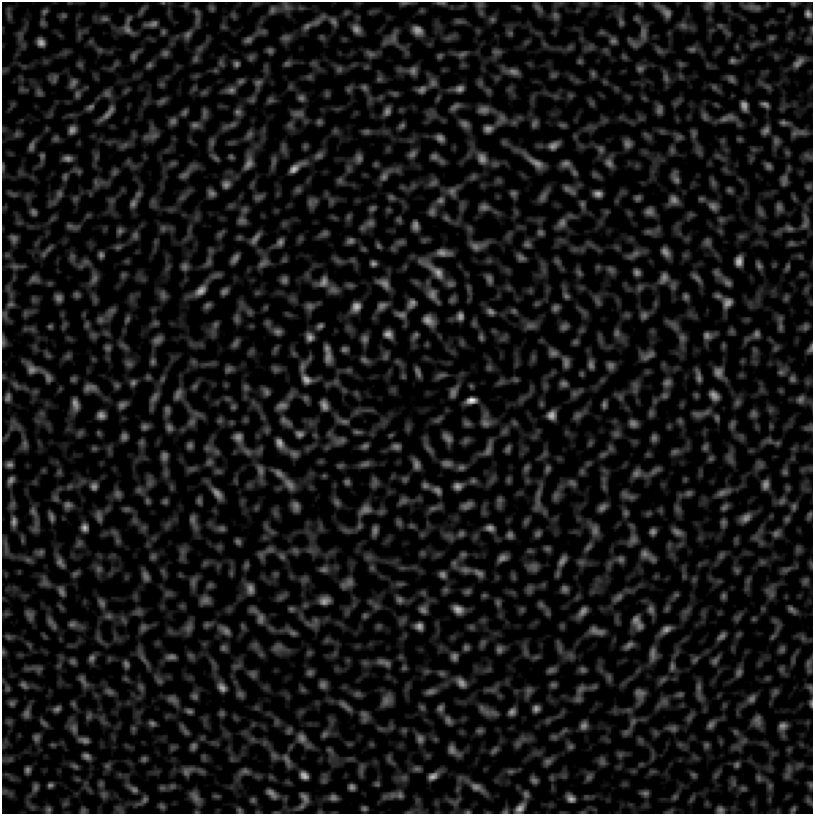}
\end{subfigure}
\begin{subfigure}
  \centering
  \includegraphics[width=.4\linewidth]{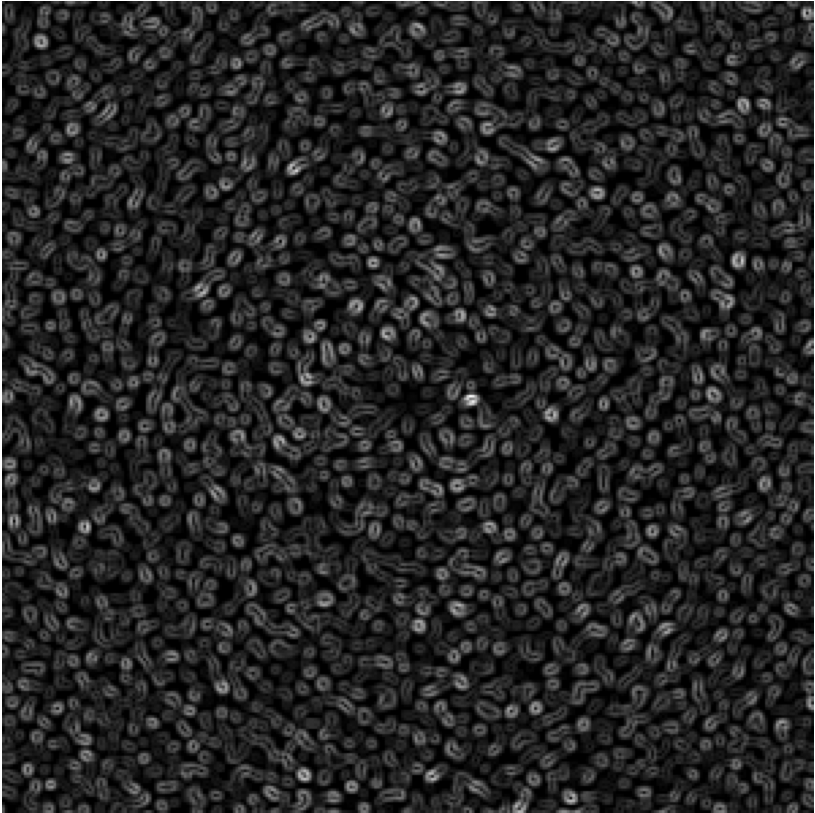}
\end{subfigure}
  \caption{An SNR map (left) and the corresponding gradient map (right).}
  \label{fig:gradmap}
\end{figure*}

The extraction of the stamps is performed using  the H2 SNR map. To reduce the number of stamps and center them on SNR peaks, we apply edge detection techniques \cite[Sobel's filter, ][]{DBLP:books/lib/DudaH73} and compute a gradient map of the SNR map. More precisely, each pixel of such a map gives an approximation of the norm of the gradient of the corresponding SNR function. An example of such a gradient map can be seen Figure \ref{fig:gradmap}. 

A pixel $u$ of an edge is identified as a pixel with a gradient value $g_u$ greater than a threshold (denoted $t$ in Algorithm \ref{alg:stampgene}). Consider the graph of \emph{edges} where vertices are pixels $u$ such that $g_u \geq t$ (edge pixels) and connect two such vertices if they are adjacent pixels in the image. More precisely, consider the graph $G=(V, E)$ with $V = \{u | g_u \geq t, u \in  \mathcal{I}\}$ and $E=\{(u,v) | u \in V, v \in V, u \textrm{ and } v \textrm{ are adjacent}\}$. An area of high SNR in the original image is related to a connected component of $G$. These components might take a variety of shapes and for sake of simplicity we consider the smallest rectangular \textbf{box} containing the component as the relevant area. 

The value of $t$ controls the number of components found and Algorithm \ref{alg:stampgene} performs a dichotomy search to reach a given average size of the boxes identifying the areas of interest. In practice we set this \emph{target} size to 12 pixels. 
Note that the stamps definition could also be done from the H3 SNR map which migh contain SNR peaks located differently. The final set of stamps would differ and might be more complete but this remains to be investigated.

\begin{algorithm}[ht!]
\caption{Stamps generation algorithm.}\label{alg:stampgene}
\hspace*{\algorithmicindent} \textbf{Input} An SNR map $\mathcal{I}$ of noise in H2, a \emph{target} size for the areas of interest\\
\hspace*{\algorithmicindent} \textbf{Output} A collection of stamps
\begin{algorithmic}[1]

\State Set the SNR of each pixel $u \in \mathcal{I}$ with a negative SNR at 0
\State Compute an approximation of the norm of the gradient $g_u$ for each pixel $u \in \mathcal{I}$
\State  $lb \leftarrow 0$, $ub \leftarrow 3\times \textrm{median}(\{g_u | u \in \mathcal{I}\})$, $t \leftarrow \frac{lb + ub}{2}$
\Do 
\State Build $G=(V, E)$ with $V = \{u | g_u \geq t, u \in  \mathcal{I}\}$ and
\Statex \indent\indent\indent $E=\{(u,v) | u \textrm{ and } v \textrm{ are adjacent pixels}\}$
\State Compute connected components of $G$ and their boxes
\If{average size of boxes $>$ \emph{target}}
\State $lb \leftarrow t$
\EndIf
\State \textbf{else } $ub \leftarrow t$ 
\State $t \leftarrow \frac{lb + ub}{2}$
\doWhile{$(ub - lb) > 10^{-3}$}
\State Build a stamp centered at the pixel of maximum SNR in each box.
\end{algorithmic}
\end{algorithm}

\subsubsection{Step (3): Training}
A simple classifier based on Logistic Regression is trained from the collection of stamps whose class is known at this stage. We outline two key points. Firstly, the classifier is dedicated to a
specific original image. The stamps generated for learning are based on injections and randomization of the original data cube on which the classifier is finally used. So that it potentially takes into account the peculiarities on this data set (e.g weather conditions). Secondly, the learning is not performed on the whole image. On the contrary, a number of limited and meaningful features are computed for each stamp. This typically refers to old-fashioned learning as opposed to deep learning. A dozen of features are computed and detailed in the Appendix. These features relate to high level descriptive statistic of the stamp such a the mean SNR value (\textbf{MeanSnr}) or the mean, maximum, and standard deviation of the gradient approximation of the stamp (\textbf{MeanGra}, \textbf{MaxGra}, \textbf{StdevGra}) but also to physical phenomenons. A key feature targets the pattern that can leave an Airy figure on the SNR map (\textbf{AiryFig}). Another one (\textbf{MeanSpec}) tries to quantify whether the 9 central pixels of a stamp indicate that the SNR signal is due to a speckle. Note that the features are not real physical models of the phenomenon but simple numerical quantities that are correlated with the presence of the phenomenon. An analysis of these correlations is presented Section \ref{featAnalysis}. Most of the features are computed on the H2 SNR  part of the stamp but a number of them are also computed on the H3 signal (typically the features dealing with the presence of a speckle). In total, a dozen of such features are used.
Finally, the logistic regression is trained using cross-validation.
\subsubsection{Step (4): Usage}
 The classifier is then used only on stamps extracted on the original image and whose center is a pixel with a SNR (in H2) greater than 2. We chose this value because below this threshold in H2, it would currently not be considered as a potential candidate by a user (an astrophysicist) in any case. A more elaborate criterion can be used, for instance using the SNR value of multiple wavelengths. But in principle, any pixel can be considered and this step is not computationally time consuming. The features of the stamp centered on that pixel are computed and if the value of the logistic regression function of the classifier is above 0.5, the stamp is kept as a candidate. It is considered to belong to the class of objects. The probability of 0.5 can be tuned and this is discussed in Section \ref{subsec:logregth}.
 
\subsubsection{Step (5): Usage - Clustering of candidates}
Since stamps centered on any pixel are potentially submitted to the classifier, adjacent pixels are likely to be classified similarly. As a result, candidates are made of close clusters that overlap and two closely overlapping stamps are most likely the same object. We therefore present the classification result as clusters of candidates.

The distance between two stamps is defined as the number of distinct pixels. The smaller this distance, the more the stamps overlap.  An agglomerative clustering algorithm \citep{mullner2011modern} merges a cluster pair if the minimum of the distances between the stamps of a cluster is lower than a threshold. In practice this threshold is set to 55.5\% of the pixels so two stamps of the same cluster overlap over at least 55.5\% (200 pixels for 19x19 stamps).

\section{Tests}

\subsection{Data sets and terminology}
\label{sec:data_terminology}

We used four IRDIS data sets obtained on HD 108767B, HIP 1993, HIP 12394 and HIP 107345. These stars were chosen as they were also used in a blind test that was performed among the SPHERE SHINE consortium to compare the merits of different algorithms (see below). They are representative of different atmospheric conditions. Table \ref{tab:targets} summarize the aforementioned stars properties and Table \ref{tab:test_star_log} provide the observing conditions.

The four data sets were used to generate \textbf{true positive} (TP) stamps (by planet's injection) and \textbf{true negative} (TN) stamps for the learning step (by processing, with algorithm \ref{alg:stampgene}, a map of pure noise generated from the original image). The blind tests consist in performing a number of injections in the original image and submitting the resulting modified data set to the classifier. Additional unknown exoplanets might also be present since the original image is used for the test.

In the following, we use the following notations and terminology: \#U is the number of stamps extracted on the original image and considered at the Usage step. The classifier is used only on stamps with a SNR greater than 2 in H2 and this number is reported in columns \#U$_{snr\geq2}$. \#TP and \#TN refers to numbers of stamps that are known to contain an injected planet or noise \emph{i.e} \emph{true positives} or \emph{true negatives}. 
Finally, when considering the results obtained with our algorithm, we note \#TPF the number of True Positives Found, and \#C the number of remaining Candidates proposed by the algorithm. These candidates are likely \textbf{false positive} but since the usage is done on the original image, it could also be real, and so far undetected, exoplanets. Therefore, we refer to them as \textbf{candidates}.

The first column of Table \ref{tab:data} reports the number of true positive (column \#TP) and negative stamps (column \#TN) obtained with our methodology and that have been used to train the classifier (column Learning). The remaining columns gives an overview of the number of stamps generated for the three Blind Tests (BT1, BT2, BT3) detailed below.
Note that some injections might lead to SNR values below 2 and thus would not be submitted to the classifier. The results are therefore given only for injected planets giving an SNR greater than 2 in H2 (the exact numbers of such injections are reported in columns titled \#TP for each blind test).

\begin{table*}
    \centering
    \begin{tabular}{|c|c|c|c|c|c|c|c|c|}
        \hline Star name & RA (J2000) & DEC (J2000) & Spectral Type & Rmag & Hmag & Age$^a$ (Myr)  \\
        \hline 
        HD108767B& 	12 29 50.8908 & -16 31 15.2081 & K 1 & 8.2 & 6.3 & $180^{+170}_{-80}$ \\  
        HIP1993 & 00 25 14.6618 & -61 30 48.2527 & M0V & 10.7 & 7.9 & $45^{+5}_{-10}$ \\  
        HIP12394 & 	02 39 35.3612 & -68 16 01.0103 & B8V& 4.1&4.4& $45^{+5}_{-10}$ \\  
        HIP107345 & 21 44 30.1227 & -60 58 38.8946 & M0 &10.5&8.1&$45^{+5}_{-10}$ \\  
        \hline 
    \end{tabular}
    \caption{Targets used for the tests.}
    \textbf{Notes.} $^a$ : Ages values extracted from \cite{2021A&A...651A..70D}
    \label{tab:targets}  
\end{table*}

\begin{table*}
    \begin{center}
        \begin{tabular}{cccccccccc}
        STAR & OBS DATE & FILTER & DIT(s)$\times$Nframe & $\Delta$PA ($\degree$)$^a$ & Seeing (")$^b$ & Airmass$^b$ & $\tau_0$ (ms)$^{a,b}$ & Program ID \\ 
        \hline 
        \hline 
        HD108767B & 2018-01-24 & DB\_H23 & 64x72 & 94.4 & 0.59 & 1.02 & 8.3 & 1100.C-0481(D) \\
        HIP1993 & 2015-11-28 & DB\_H23 & 64x64 & 25.8 & 1.57 & 1.26 & 7.2 & 096.C-0241(B) \\
        HIP12394 & 2016-09-15 & DB\_H23 & 32x160 & 29.1 & 0.42 & 1.38 & 9.2 & 097.C-0865(D)  \\
        HIP107345 & 2015-07-04 & DB\_H23 & 64x64 & 26.0 & 1.07 & 1.25 & 2 & 095.C-0298(C) \\       
        \hline 
        \hline        
        \end{tabular}    
    \caption{Test targets observation logs. \textbf{Notes:} $^a$: DIT corresponds to the detector integration time per frame, $\Delta$PA is the amplitude of the parallactic rotation, $\tau_0$ corresponds to the coherence time. $^b$: Values extracted from the updated DIMM info and averaged over the sequence. }
    \end{center}
    \label{tab:test_star_log}
\end{table*}

\begin{figure*}
\begin{subfigure}
  \centering
  \includegraphics[width=.3\linewidth]{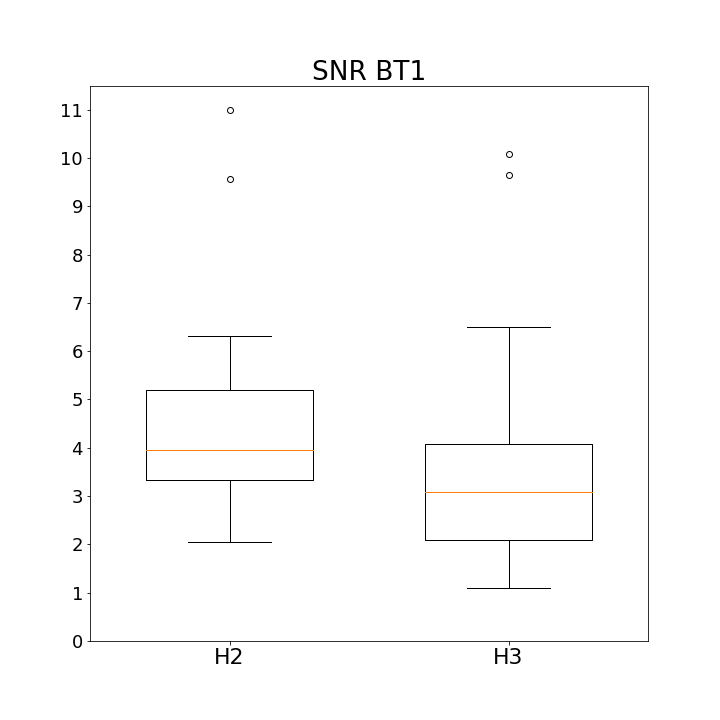}
\end{subfigure}%
\begin{subfigure}
  \centering
  \includegraphics[width=.3\linewidth]{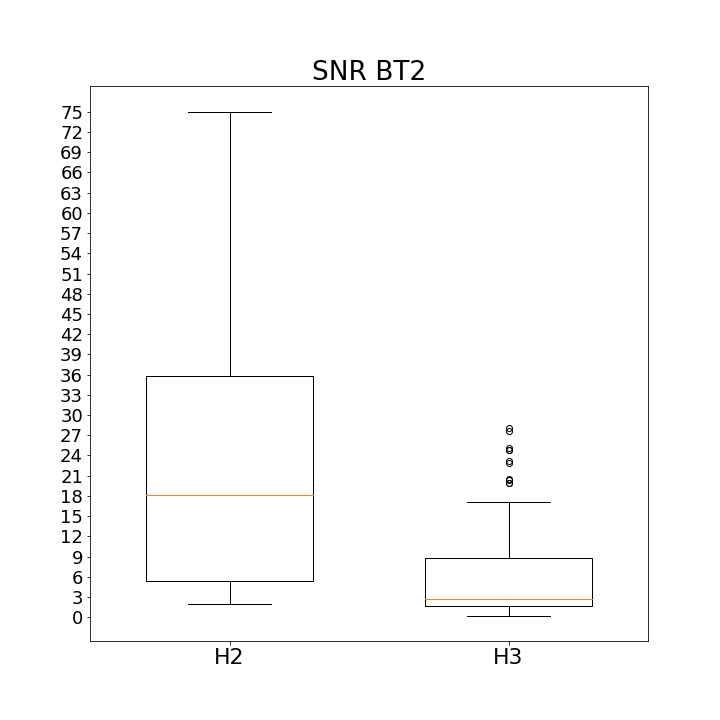}
\end{subfigure}
\begin{subfigure}
  \centering
  \includegraphics[width=.3\linewidth]{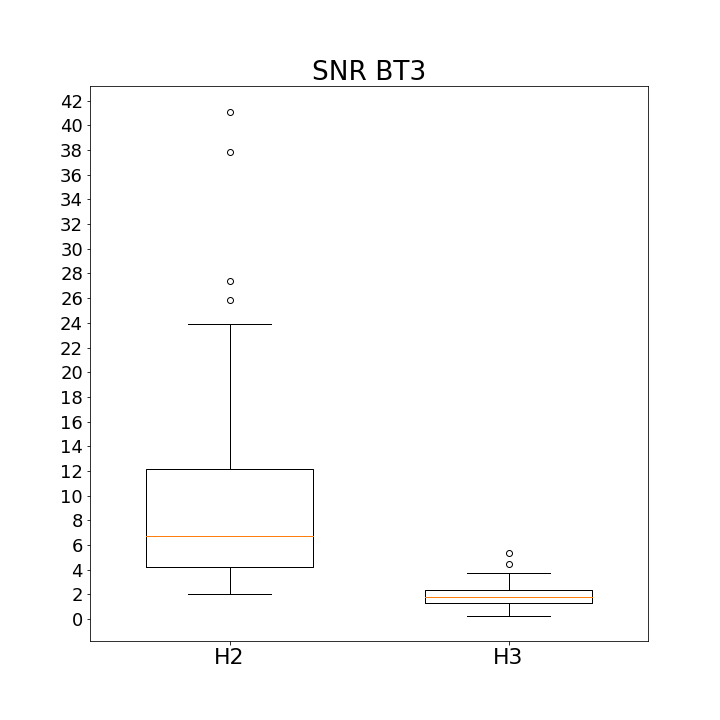}
\end{subfigure}
\caption{Box plots summarizing the SNR distribution of the injections in H2 and H3 (resp. left and right in each boxplot) for the three Blind Tests: BT1, BT2, BT3. Each boxplot shows the minimum, first quartile, median, third quartile, maximum (by convention, the third quartile plus at most 1.5 of the inter-quartile range) and additional points above.}
\label{fig:snrdistrib}
\end{figure*}

\subsection{Tests description}



We started with a blind test constructed by the SHINE consortium, BT1. Eight fake companions featuring spectral types between early M and late T were injected for each of the stars considered; their contrasts were chosen to provide a TLOCI signal \citep{MAROIS_TLOCI} of about 5 sigma (mean of H2 and H3 contrasts). We note that the objects do not necessarily represent realistic planets. 
Note also that in a few cases, the injected planets are not detected with PACO even with a low threshold. This happens when the planets are very close to the stars, and their expected  contrasts are overestimated. 
Finally, one out of the 8 planets around  HD10767B has an SNR in H2 lower than 2, and two out of the 8 planets around HIP12394 have to SNR in H2 lower than 2. They are therefore not be considered here.

BT2 and BT3 are blind tests generated using the process described in section \ref{sec:injection}. The planet's masses considered are 1, 2, 3, 4 and 5 MJup. The contrasts of the companions wrt the stars are computed according to their masses and to the age of the stars (planets and stars are assumed to be coeval), and using the COND models. Conversely to BT1, the fake planets correspond to realistic cases. We note that the respective contrasts in BT1 are very different from those in BT2 and BT3.
 Figure \ref{fig:snrdistrib} shows the distribution of the SNR values of the injections in the three blind tests.

The last test, BT4, is similar to BT2 and BT3. We massively injected FPs close to the star, between 0.25 and 1 arcsec. The FPs were chosen so that their contrasts be along the 5sigma contrast curves to test the capability of the classifier. We chose to use the classifier on HIP12394 with 3Mjup and 4Mjup injected planets because both planets cross the contrast curve between 0 and 1 arcsec.

\begin{table*}
    \centering
    \begin{tabular}{|c|ccc|cc|cc|cc|}
        \hline star name &  \multicolumn{3}{|c|}{Learning} & \multicolumn{2}{|c|}{BT1} & \multicolumn{2}{|c|}{BT2} & \multicolumn{2}{|c|}{BT3}\\ \hline 
        & \#TP & \#TN & \#TN$_{snr\ge 2}$ & \#TP & \#U$_{snr\ge 2}$ & \#TP & \#U$_{snr\ge 2}$ & \#TP & \#U$_{snr\ge 2}$\\ \hline 
        all&1026 & 77565 & 8180 & 29 & 14691 & 140 & 19017 & 90 & 15878\\  \hline 
        HD108767B&259 & 15621 & 1724 & 7 & 3577 & 26 & 3777& 15 & 3390 \\  \hline 
        HIP1993&253 & 15228 & 1271 & 8 & 2376 & 37 & 3479 & 24 & 2810 \\  \hline 
        HIP12394&251 & 15636 & 2920 & 6 & 6670 & 39 & 8226 & 25 & 7191\\  \hline 
        HIP107345&263 & 31080 & 2265 & 8 & 2068 & 38 & 3535 & 26 & 2487\\  \hline 
    \end{tabular}
    \vspace{0.4cm}
    \caption{Number of positive (planet's injection) and negative stamps (extracted from a pure noise map generated with algorithm \ref{alg:stampgene}). The total number as well as the number of negative stamps of SNR H2 greater than 2 are reported. \label{tab:data}}  
\end{table*}

\begin{table*}
    \centering
    \begin{tabular}{|c|cc|cc|}
        \hline 
        star name &  \multicolumn{2}{|c|}{Learning} & \multicolumn{2}{|c|}{BT4}\\ \hline 
        & \#TP & \#TN & \#TP & \#U$_{snr\ge 2}$\\ \hline 
        HIP1993&243 & 15076 & 34 & 2863  \\  \hline 
        HIP12394-4M&214 & 15540 & 34 & 8384  \\  \hline 
        HIP12394-3M&214 & 15540 & 34 & 7707  \\  \hline 
        HIP107345&192 & 15791 & 34& 2421 \\  \hline 
    \end{tabular}
    \vspace{0.4cm}
    \caption{Number of positive (planet's injection) and negative stamps (extracted from a pure noise map generated with algorithm \ref{alg:stampgene}). The total number as well as the number of negative stamps of SNR H2 greater than 2 are reported. \label{tab:data}}  
\end{table*}

\subsection{Features analysis}
\label{featAnalysis}
\begin{figure*}
\begin{subfigure}
  \centering
  \includegraphics[width=.32\linewidth]{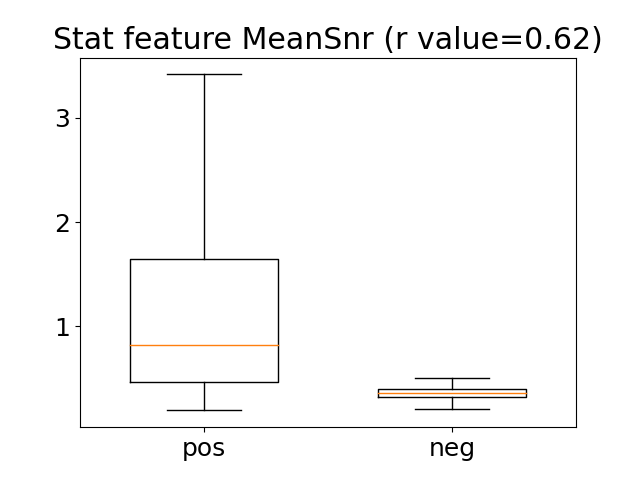}
\end{subfigure}%
\begin{subfigure}
  \centering
  \includegraphics[width=.32\linewidth]{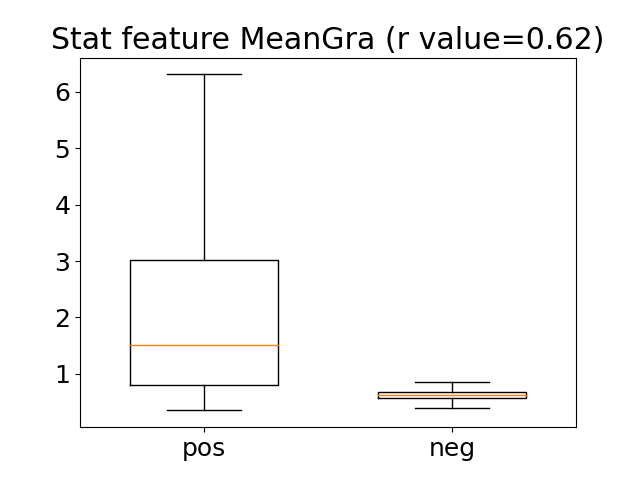}
\end{subfigure}
\begin{subfigure}
  \centering
  \includegraphics[width=.32\linewidth]{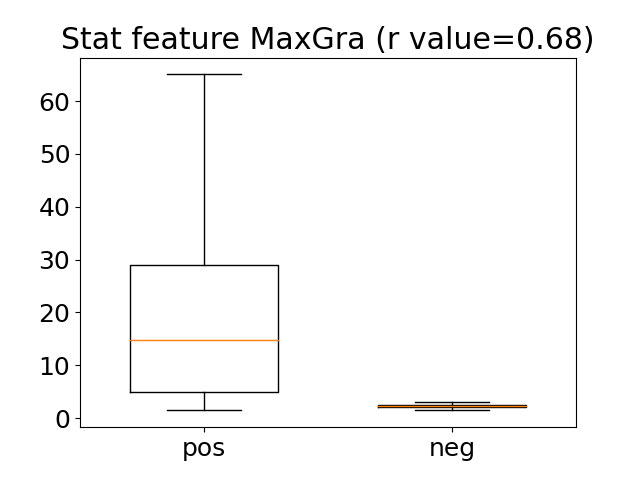}
\end{subfigure}
\begin{subfigure}
  \centering
  \includegraphics[width=.32\linewidth]{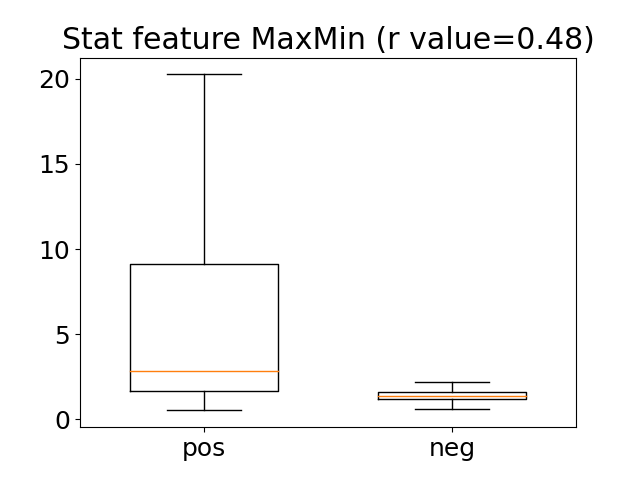}
\end{subfigure}%
\begin{subfigure}
  \centering
  \includegraphics[width=.32\linewidth]{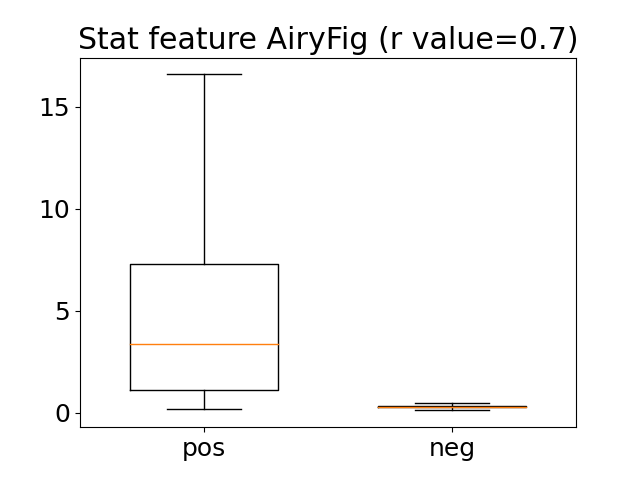}
\end{subfigure}
\begin{subfigure}
  \centering
  \includegraphics[width=.32\linewidth]{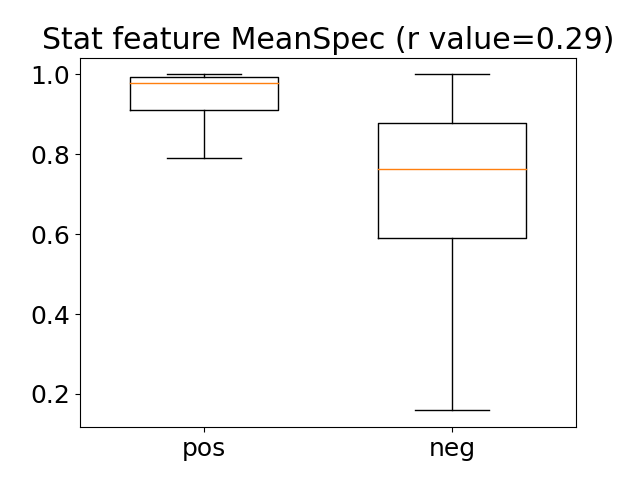}
\end{subfigure}
\caption{Box plots summarizing the distribution of six features by showing the minimum, first quartile, median, third quartile, maximum. The distribution for positive and negative stamps are shown for each feature to help visualize whether the feature discriminates the two classes. The values are computed over stamps of SNR H2 $\geq 2$ of the four images (In total 2069 positives, 18074 negatives).}
\label{fig:featureanalysis}
\end{figure*}

We analyse the distribution of six of the features (\textbf{MeanSnr}, \textbf{MeanGra}, \textbf{MaxGra}, \textbf{MaxMin}, \textbf{AiryFig}, \textbf{MeanSpec}) across true positive and true negative stamps. We recall that a feature is simply a real number computed on a stamp and a good feature is correlated to a class (positive/negative). Figure \ref{fig:featureanalysis} provides,  for each feature, two box plots showing the distribution of its values for true positive (box labeled pos) and true negative (box labeled neg) stamps. A box plot gives a summary of the distribution in 5 numbers from bottom to top: minimum, first quartile, median, third quartile and maximum. Typically, the values of the feature for 50\% of the stamps lie in the box. The median value is the orange horizontal line within the box. Values considered as outliers (below or above 1.5 the inter-quartile range, which is defined as the maximum) are not shown for sake of clarity. The correlation coefficient (r value) is given for each feature. 

We expect useful features to show distinct distributions for positive and negative stamps in order to help the classifier discriminating between the two. The size of the intersection of the two distributions (for the positive and negative class) gives an idea of the discrimination power of the feature. 

Four features appear decisive (correlation coefficient $r \geq 0.6$): the mean snr intensity (\textbf{MeanSnr}), the gradient features (\textbf{MeanGra}, \textbf{MaxGra}) as well as the feature related to the presence of an Airy Figure (\textbf{AiryFig}).

\begin{figure*}
\begin{subfigure}
  \centering
  \includegraphics[width=.3\linewidth]{{boxplot_snr_min2.0_dist_min0_dist_max2000_all_MeanSpec}.png}
\end{subfigure}%
\begin{subfigure}
  \centering
  \includegraphics[width=.3\linewidth]{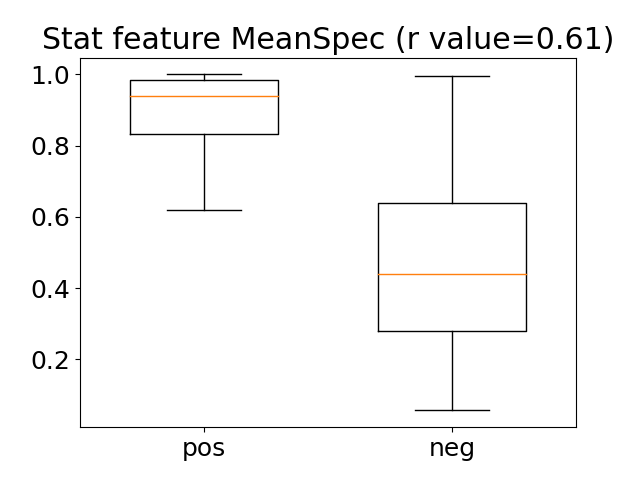}
\end{subfigure}
\begin{subfigure}
  \centering
  \includegraphics[width=.3\linewidth]{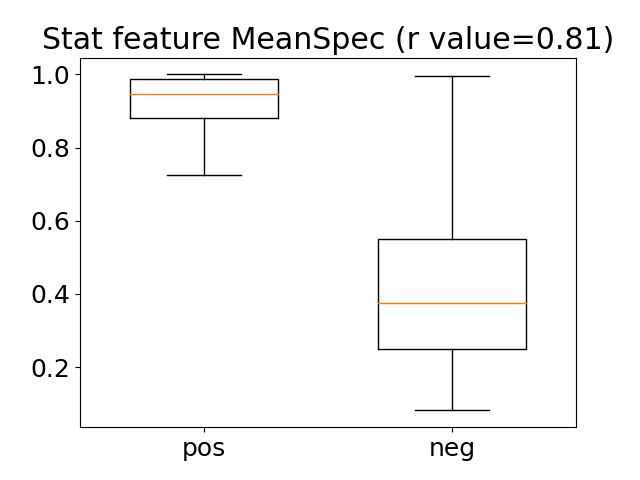}
\end{subfigure}
\caption{Distribution of the \textbf{MeanSpec} feature for three selected subsets of stamps respectively from left to right: (1) all stamps of SNR H2 $\geq 2$ (2069 positives, 18074 negatives),  (2) all stamps of SNR H2 $\geq 2$ located between 30 and 140 pixels (554 positives, 1969 negatives),  (3) all stamps of SNR H2 $\geq 2.5$ located between 30 and 140 pixels  (492 positives, 558 negatives).}
\label{fig:distribution_speckle}
\end{figure*}

The feature related to speckles is not strongly correlated to the presence of a companion ($r = 0.29$). This is expected as speckles are only present within the star halo. Restricting the analysis to the halo region, between 30 and 140 pixels (i.e. 370 to 1700 mas), significantly increases the correlation to $r = 0.61$ (see Figure \ref{fig:distribution_speckle}). It might therefore be appropriate to build two distinct classifiers, one for the star halo that includes \textbf{MeanSpec} and one for the remaining area without it. But this tends to complicate further the overall process. We decided to keep it simple for the moment and used the feature \textbf{MeanSpec} on the entire field with an additional 0/1 indicative feature ($f_9$ in the Appendix) defining whether a stamp is or not in the star halo. In other words, this indicative feature tells when \textbf{MeanSpec} is relevant and can be eventually help the classifier.  

\subsection{Analysis}
\label{analysis}

The results are presented in Table \ref{tab:totres} for BT1, BT2, BT3 and BT4. The results of the classifier (column RegL) are compared to the threshold approach (thresholds of 3 and 5) for two flavors of PACO: ADI and ASDI. In ADI, the SNR is computed considering  separately the H2 and H3 channels. So a detection is made if any of the SNR value in H2 or H3 is above the threshold. In ASDI, PACO optimally combines both bands  \cite{flasseur2020ASDI}. The number of planets found (\#TPF) and the number of candidates (\#C) are reported for each approach (see Section \ref{sec:data_terminology} for the definition of \#TPF and \#C).

Overall, ADI, ASDI 3 and RegL retrieve most of the planets. Yet, the former two detect also a huge number of candidates: typically several hundreds candidates are found in ASDI 3 and often about one hundred in ADI 3, while RegL detects a much more limited number of candidates. Such large lists of candidates found by ADI and ASDI 3 prevent from identifying the true positives. This justifies the usual choice of a threshold of 5 when using PACO. Using a threshold of 5 removes nearly all candidates but the detection performances are reduced (in particular for BT2 but also BT3) compared to our classifier. Hence, RegL performs better than ADI or ASDI, giving a better compromise between the number of detections and the number of candidates.
We detail the results of each test below.



\begin{table*}
\centering
\begin{tabular}{l|ccccccccccc|}
\cline{2-12}
                                & \multicolumn{1}{c|}{}       & \multicolumn{2}{c|}{RegL}                   & \multicolumn{2}{c|}{ADI 3}                      & \multicolumn{2}{c|}{ADI 5}                      & \multicolumn{2}{c|}{ASDI 3}                        & \multicolumn{2}{c|}{ASDI 5}                        \\ \cline{2-12} 
                                & \multicolumn{1}{c|}{\#Inj} & \multicolumn{1}{c|}{\#TPF} & \multicolumn{1}{c|}{\#C} & \multicolumn{1}{c|}{\#TPF} & \multicolumn{1}{c|}{\#C} & \multicolumn{1}{c|}{\#TPF} & \multicolumn{1}{c|}{\#C} & \multicolumn{1}{c|}{\#TPF} & \multicolumn{1}{c|}{\#C} & \multicolumn{1}{c|}{\#TPF} & \multicolumn{1}{c|}{\#C} \\ 
                                \cline{2-12}
                                
                                & \multicolumn{11}{c|}{\textbf{BT1}}                                                                                                                                                                                                                                                                                                                    \\ \hline
\multicolumn{1}{|l|}{HD108767B} & \multicolumn{1}{r|}{7}      & 5          & \multicolumn{1}{r|}{11}     & 6            & \multicolumn{1}{r|}{153}    & 3    & \multicolumn{1}{r|}{0}      & 6  & \multicolumn{1}{r|}{717}    & 5   & 0    \\
\multicolumn{1}{|l|}{HIP1993}   & \multicolumn{1}{r|}{8}      & 6          & \multicolumn{1}{r|}{1}      & 8            & \multicolumn{1}{r|}{100}     & 3    & \multicolumn{1}{r|}{0}      & 8   & \multicolumn{1}{r|}{724}    & 7   & 0   \\
\multicolumn{1}{|l|}{HIP12394}  & \multicolumn{1}{r|}{6}      & 5          & \multicolumn{1}{r|}{2}      & 5            & \multicolumn{1}{r|}{397}    & 2    & \multicolumn{1}{r|}{0}      & 6   & \multicolumn{1}{r|}{611}    &  5  & 0   \\
\multicolumn{1}{|l|}{HIP107345} & \multicolumn{1}{r|}{8}      & 7          & \multicolumn{1}{r|}{2}      & 8            & \multicolumn{1}{r|}{65}     & 2    & \multicolumn{1}{r|}{0}      & 8  & \multicolumn{1}{r|}{694}    & 8      & 0  \\ \hline
\multicolumn{1}{|l|}{Total}     & \multicolumn{1}{r|}{29}     & 23         & \multicolumn{1}{r|}{16}     & 27           & \multicolumn{1}{r|}{715}    & 10   & \multicolumn{1}{r|}{0}      & 28  & \multicolumn{1}{r|}{2746}   & 25  & 0    \\ \hline

                                & \multicolumn{11}{c|}{\textbf{BT2}}                                                                                                                                                                                 \\ \hline
\multicolumn{1}{|l|}{HD108767B} & \multicolumn{1}{r|}{26}     & 22          & \multicolumn{1}{r|}{13}     & 21          & \multicolumn{1}{r|}{2}     & 16   & \multicolumn{1}{r|}{0}      & 23  & \multicolumn{1}{r|}{716}    & 18   & 1    \\
\multicolumn{1}{|l|}{HIP1993}   & \multicolumn{1}{r|}{37}     & 33          & \multicolumn{1}{r|}{1}      & 32          & \multicolumn{1}{r|}{91}    & 27   & \multicolumn{1}{r|}{0}      & 31  & \multicolumn{1}{r|}{627}    & 23   & 1    \\
\multicolumn{1}{|l|}{HIP12394}  & \multicolumn{1}{r|}{39}     & 27          & \multicolumn{1}{r|}{1}      & 28          & \multicolumn{1}{r|}{373}   & 20   & \multicolumn{1}{r|}{0}      & 24  & \multicolumn{1}{r|}{620}    & 22   & 1    \\
\multicolumn{1}{|l|}{HIP107345} & \multicolumn{1}{r|}{38}     & 38          & \multicolumn{1}{r|}{1}      & 38          & \multicolumn{1}{r|}{77}    & 33   & \multicolumn{1}{r|}{0}      & 36  & \multicolumn{1}{r|}{659}    & 33   & 1    \\ \hline
\multicolumn{1}{|l|}{Total}     & \multicolumn{1}{r|}{140}    & 120         & \multicolumn{1}{r|}{16}     & 119         & \multicolumn{1}{r|}{543}   & 96   & \multicolumn{1}{r|}{0}      & 114  & \multicolumn{1}{r|}{2622}   & 96   & 4    \\ \hline

 & \multicolumn{11}{c|}{\textbf{BT3}}                                                                                                                                                                                                                                                                                                                    \\ \hline
\multicolumn{1}{|l|}{HD108767B} & \multicolumn{1}{r|}{15}      & 6          & \multicolumn{1}{r|}{10}     & 8            & \multicolumn{1}{r|}{118}    & 2     & \multicolumn{1}{r|}{0}      & 8  & \multicolumn{1}{r|}{440}    & 3    & 1    \\
\multicolumn{1}{|l|}{HIP1993}   & \multicolumn{1}{r|}{24}      & 20         & \multicolumn{1}{r|}{4}     & 19           & \multicolumn{1}{r|}{65}     & 19    & \multicolumn{1}{r|}{0}      & 23  & \multicolumn{1}{r|}{390}  & 18   & 0   \\
\multicolumn{1}{|l|}{HIP12394}  & \multicolumn{1}{r|}{25}      & 19         & \multicolumn{1}{r|}{2}     & 19           & \multicolumn{1}{r|}{348}    & 16    & \multicolumn{1}{r|}{0}      & 20 & \multicolumn{1}{r|}{386}    & 19   & 1    \\
\multicolumn{1}{|l|}{HIP107345} & \multicolumn{1}{r|}{26}      & 24         & \multicolumn{1}{r|}{2}     & 24           & \multicolumn{1}{r|}{61}     & 21    & \multicolumn{1}{r|}{0}      & 26 & \multicolumn{1}{r|}{414}    & 24   & 1    \\ \hline
\multicolumn{1}{|l|}{Total}     & \multicolumn{1}{r|}{90}      & 69         & \multicolumn{1}{r|}{18}     & 70           & \multicolumn{1}{r|}{592}    & 58    & \multicolumn{1}{r|}{0}     & 77 & \multicolumn{1}{r|}{1630}   & 64   & 3   \\ \hline

& \multicolumn{11}{c|}{\textbf{BT4}}                                                                                                                                                                                                                                                                                                                    \\ \hline

\multicolumn{1}{|l|}{HIP1993}   & \multicolumn{1}{r|}{34}      & 30          & \multicolumn{1}{r|}{7}      &      31       & \multicolumn{1}{r|}{100}     &  19   & \multicolumn{1}{r|}{2}      &  32  & \multicolumn{1}{r|}{956}    &  28  &  1  \\
\multicolumn{1}{|l|}{HIP12394-4MJup}  & \multicolumn{1}{r|}{34}      & 33          & \multicolumn{1}{r|}{12}      &    34       & \multicolumn{1}{r|}{392}    &  26  & \multicolumn{1}{r|}{7}      &  33  & \multicolumn{1}{r|}{909}    &  29  &  0  \\
\multicolumn{1}{|l|}{HIP12394-3MJup}  & \multicolumn{1}{r|}{34}      & 21          & \multicolumn{1}{r|}{12}      &    25       & \multicolumn{1}{r|}{388}    &  12  & \multicolumn{1}{r|}{7}      &  26  & \multicolumn{1}{r|}{734}    &  13  &  0  \\
\multicolumn{1}{|l|}{HIP107345} & \multicolumn{1}{r|}{34}      & 27          & \multicolumn{1}{r|}{2}      &     29        & \multicolumn{1}{r|}{75}     &  18   & \multicolumn{1}{r|}{2}      & 30  & \multicolumn{1}{r|}{951}    &  30     &  0 \\ \hline
\multicolumn{1}{|l|}{Total}     & \multicolumn{1}{r|}{136}     &  111        & \multicolumn{1}{r|}{33}     &     119       & \multicolumn{1}{r|}{955}    & 75   & \multicolumn{1}{r|}{18}      &  121 &  \multicolumn{1}{r|}{3550}  &  100 & 1    \\ \hline

\end{tabular}
\vspace{0.3cm}
\caption{Comparison between the results from the logistic regression approach (Column RegL) and  those from threshold detections. Two thresholds are considered: 3 and 5, and two setups are considered: ADI and ASDI. Columns $\#$TPF reports the number of planets found whereas $\#$C is the number of additional candidates produced. $\#$Inj is the number of injections (hidden planets) in the corresponding blind test.\label{tab:totres}}
\end{table*}

BT1. RegL finds 23 planets while ADI 3 finds 27. Yet, RegL finds 16 false positives, to be compared to the 715 found by ADI 3. 
ASDI 5 find 25 planets and avoids any additional candidates. We will discuss this result below. 

BT2. RegL retrieves 120  out of the 140 injected planets, identifying one more planet than ADI 3. Noticeably, RegL finds only 16 additional candidates (likely false positive), to be compared to the 543 found with the threshold approach. ASDI 3 finds 114 planets and produces much more additional candidates than RegL. RegL finds more planets than ADI and ASDI 5 (which find 96 planets), with a yet slightly larger number of candidates (16 instead of 0 and 4, respectively).

BT3. RegL does not performs better than ADI or ASDI 3 regarding the number of planets found but the number of candidates with RegL (18) is significantly smaller with with ADI (592) or ASDI (1630). It behaves better when compared to ADI or ASDI 5, with a larger amount of detected planets (69 versus 58 with ADI 5 and 64 with ASDI 5), and a small number of candidates (18 versus  0 and 3, respectively).

BT4. RegL does not perform better than ADI or ASDI 3 regarding the number of planets found but the number of candidates with RegL is much smaller than with ADI 3 or ASDI 3. It is relevant to notice that despite of the little gap between the number TPF and the number of planets injected, RegL finds for HIP1993, HIP12394-4M and HIP107345 almost all the sources with a center SNR>2 and misses only 5 planets with SNR very close to 2 for HIP12394-3M. In fact, only 26 planets out of the 34 have an SNR above the threshold of RegL. This gap can be explained by the fact that we injected along the 5sigma contrast curves at the constant mass so some planets close to the star are below this curve.

\begin{figure}
\centering
  \includegraphics[width=\linewidth]{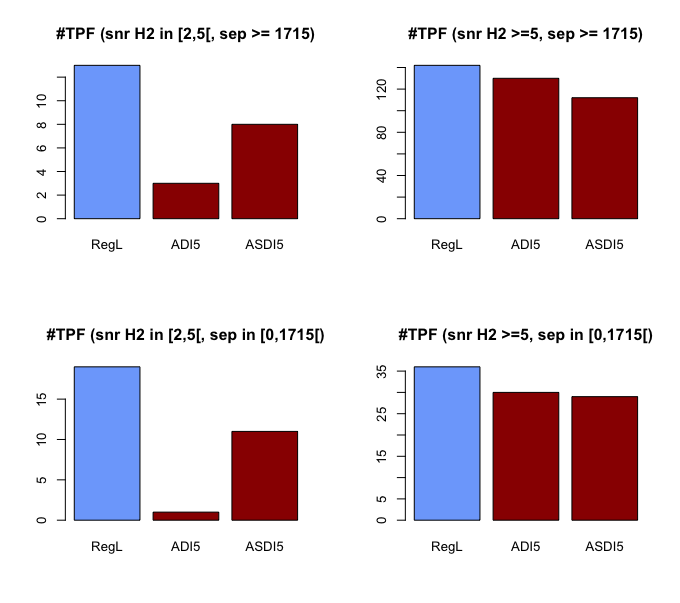}
  \caption{Number of detections (\#TPF) achieved by the different methods depending on the SNR and separation from the stars of the targets.\label{sepSNRquadrants}}
  \label{fig:speckle}
\end{figure}

Figure \ref{sepSNRquadrants} shows the number of planets found ($\#$TPF) (for all blind tests) depending on their projected separations to the star (in $[0,1715[$ or $\geq 1715$ mas), as well as on their SNR value \textbf{in H2} (in $[2,5[$ or $\geq 5$). Results for ADI 3 are not reported here since they lead to far too many candidates. As expected, the logistic regression improves over ADI/ASDI for low H2 SNR \footnote{A detection can be made by ASDI 5 below an SNR threshold of 5 in H2 because PACO combines both H2 and H3 bands in ASDI. It can also be made in ADI 5 if the SNR in H3 alone is above 5.}. It can also be noted here that most of the injections of these blind tests have a high SNR and are located outside of the star halo.

Finally, note that the classifier is expected to perform better on BT2/BT3 than on BT1 because the exact same injection process is used at the learning step and the usage step for BT2/BT3. Conversely, the injections of BT1 were performed independently of the present work with different parameters (thus a different opinion on what is most realistic) and can show different patterns that are therefore not learnt by the classifier. In particular, injections of BT1 can have a low SNR in H2 but a high a SNR in H3. This pattern is not typical of the injection process used in the present work (section \ref{sec:injection}) for the learning step. As a result, the classifier does not learn to consider such patterns and is less efficient on BT1. It remains competitive but its performances on BT1 could certainly be improved by including injections at the learning step that are consistent with the one used for the test.






\subsection{Image dedicated classifier versus a single classifier.} 
\label{singleCL}
So far in our study, a classifier is learnt for each image and its usage is dedicated to the original image from which the exoplanets are to be detected. This has the advantage to take into account peculiarities of the image such as the weather conditions at the time it was taken. The alternative is to train a single and common classifier using the four data sets together aiming for better generalization.

The table of Figure \ref{tab:SingleVSAll} gives the result of such a single classifier (first line) trained over all data sets. The two approaches produce very similar results. The single classifier identifies one additional planet on BT3 (70 versus 69) but misses one on BT1 (22 versus 23) and tend to produce the same amount of candidates that are not injected planets. However, these results might considerably depend on the quality and uniformity of the images. We expect that a larger data set sampling a wider range of weather conditions and quality might be needed to investigate this option further.

\begin{figure*}																			\centering
\begin{tabular}{|c|c|cc|c|cc|c|cc|}			\hline																			
	&	\multicolumn{3}{|c|}{	BT1	} &	 \multicolumn{3}{|c|}{	BT2	}  &	 \multicolumn{3}{|c|}{	BT3	}\\
	\hline																					
        	&\#Inj	&\#TPF&\#C	&	\#Inj	&	\#TPF	&	\#C	&	\#Inj	&	\#TPF	&	\#C	\\
	\hline																					
	Single classifier	    &		&		&		&		&		&	&		&		&		\\
	\hline
    4 stars 	&	29	&	22	&	15	& 140	    &	\textbf{120}	&	13 & 90		&	\textbf{70}	&	16\\
	\hline	\hline																			
	One classifier	    &		&		&		&		&		& &		&		&\\
	per image	    &		&		&		&		&		& &		&		&\\
	\hline
	HD108767B	&	7	&	5	&	11	& 26	&	22	&	13	& 15	& 6	    &	10	\\
	HIP1993	    &	8	&	6	&	1	& 37	&	33	&	1	& 24	& 20	&	4	\\
	HIP12394	&	6	&	5	&	2	& 39	&	26	&	1	& 25	& 19	&	2	\\
	HIP107345	&	8	&	7	&	2   & 38	&	38	&	1   & 26	& 24	&	2	\\
	\hline
	Total	    &   29	&	\textbf{23}	&	16	&	140	&	\textbf{120}	&	16	& 90& 	69	&	18	\\
	\hline	\hline
  Two classifiers	    &		&		&		&		&		&	&		&		&		\\
	\hline
    4 stars 	&	29	&	19	&	\textbf{14}	& 140	    &	119	&	\textbf{6} & 90		&	\textbf{70}	&	\textbf{14}\\
	\hline																		
\end{tabular}																		
\vspace{0.3cm}
\caption{Comparing a single classifier trained over the 4 data sets to a classifier dedicated to each data set.\label{tab:SingleVSAll}}
\end{figure*}				
    
\subsection{Handling speckles with two classifiers.} 

A speckle moves radially away from the center between the two wavelengths depending on its distance to the center and the ratio $r = \frac{\lambda_{H3}}{\lambda_{H2}}$. One of the feature proposed tries to take advantage of this motion to detect speckles. But since the motion depends on the distance to the star, a speckle near the center might shift less than a single pixel which makes the feature irrelevant. Additionally, speckles are not present far from the star where another regime occurs and the noise tends to be dominated by photon and instrument noise. Overall, the feature is only valid at a minimum $d_{min}$ and maximum $d_{max}$ distance from the star and associated to an indicator (another feature) that  is set to one when the center of the stamp is located in the range $[d_{min}, d_{max}]$ (see the Annexe for details). \\
 
Another approach is to train two distinct classifiers. The first one is trained with stamps included in  $[d_{min}, d_{max}]$ and includes the speckle feature whereas the second one is trained with stamps within $[0, d_{min}[ \cup ]d_{max}, +\infty[$ and does not use the speckle feature. The results obtained with two classifiers are shown in the last row of Figure \ref{tab:SingleVSAll} and appear to be very similar or slightly worse. Additional training samples might be required to properly train two classifiers as opposed of one.
    
\subsection{Threshold of the logistic regression.}
\label{subsec:logregth}

Logistic regression gives a probability of belonging to a class. By default, a threshold of 0.5 is used to decide whether the object belongs to the class or not. However, it is possible to choose another threshold value and thus obtain more or less samples classified as true.
A classical way to determine this threshold is to use a ROC curve (receiver operating characteristic), the true positive rate against the false positive rate. But, in the case of imbalanced data, the precision recall curve is often preferred.
Precision is the proportion of relevant items among all the proposed items; recall is the proportion of relevant items proposed among all the relevant items.

To determine a threshold from precision and recall, we use the f-score calculated as follows:
$F_\beta = (1 + \beta^2) \cdot \frac{\mathrm{precision} \cdot \mathrm{recall}}{(\beta^2 \cdot \mathrm{precision}) + \mathrm{recall}}$

\begin{figure}
  \centering
  \includegraphics[width=\linewidth]{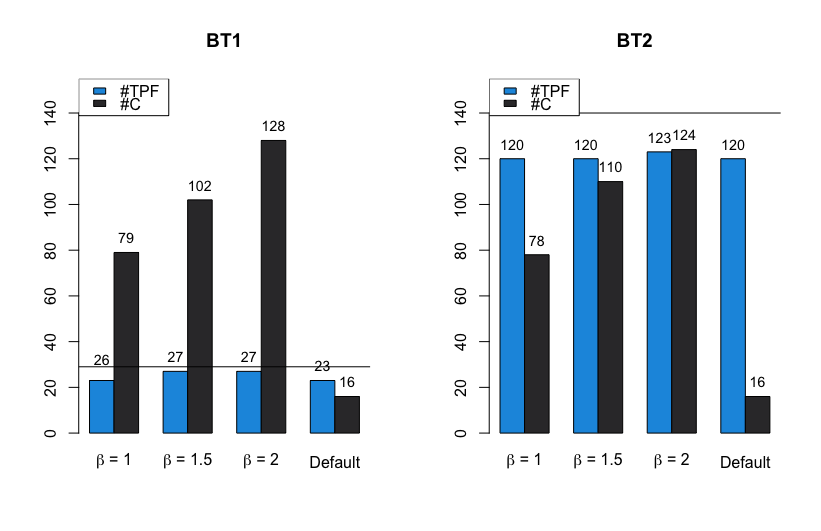}
\caption{Number of objects found ($\#$TPF in blue) and candidates ($\#$C in black) according to the value of $\beta$. The black line represents the number of injections (29 for BT1 and 140 for BT2).}
\label{fig:fscore}
\end{figure}

Increasing the value of $\beta$ increases the weight of the precision. The threshold that maximizes the f-score on the training data is calculated and then used for classification. As can be seen in figure \ref{fig:fscore}, as $\beta$ increases, the number of candidates proposed can increase, as well as the number of objects found. Comparing it with a threshold set at $0.5$ (on the right, Figure \ref{fig:fscore}), we find a larger number of objects, but also a much greater number of candidates.

The classifier can therefore be adjusted to the best compromise (number of true positives versus number of false positives) that suits the user. It depends on how much effort a user can afford to check by hand the candidates in order to increase the chance of a true detection. In the present case, we note that a significant increase of the number of candidates is required to detect only a few more exoplanets.


\subsection{Amount of data required.} We evaluated the amount of training data required to reach the performance reported for the single classifier. The injection process can require a non negligible computational effort in practice. It turns out to be more costly than the learning, clustering and classification steps. To evaluate the real need for the injections, we run the training for the following numbers of injected planets: \#pos  $\in \{25,50,100,150,200,250\}$. We also investigate different number of guaranteed noise stamps with \#neg $ \in \{1000, 5000, 10000, 15000\}$ to get a sense of the effect of imbalance. 
This analysis was restricted to BT2.


 Overall the 120 planets (actual performance of our approach on BT2) are found using only 150 injected planets in the training stage (whereas 250 injections were initially used). A minimum of 10000 negative samples are needed to avoid too many false positives. 

\section{Application to 51 Eri}
\subsection{51 Eridani}
51 Eridani (HIP 21547) is an F0-type star that hosts one 2-4 $\text{M}_{\text{jup}}$ planet imaged in 2014 by the Gemini Planet Imager  \citep{Macintosh2015_51eri}. We apply the proposed methodology to four IRDIS data sets  taken with IRDIS on 25/12/2015, 15/01/2016, 11/12/2016 and 12/12/2016. The observing log and setup of these 4 observations can be found in Table. \ref{tab:51eri_log}. Note that on Dec 2016, the data, obtained under good atmospheric conditions, are affected by the so called low wind effect that occurred when the wind was very low, and considerably degraded the image quality \citep{milli2018}\footnote{this effect has been taken care of afterwards}. 

We consider the four data sets completely independently, as our approach is not informed of the temporality. Table \ref{tab:eridani} reports the SNR of the planet in each image with PACO ADI and PACO ASDI (column SNR), whether it was found or not by our method (yes/no of column Found) and the number of candidates proposed (column \#C). 

\begin{table*}
    \begin{center}
        \begin{tabular}{cccccccccc}
        STAR & DATE OBS & FILTER & DIT(s)$\times$Nframe & $\Delta$PA ($\degree$)$^a$ & Seeing (")$^b$ & Airmass$^b$ & $\tau_0$ (ms)$^{a,b}$ & Program ID \\ 
        \hline 
        \hline 
        HIP 21547 & 2015-12-25 & DB\_H23 & 16x256 & 37.6 & 1.18 & 1.10 & 1.8 & 096.C-0241(C) \\
        HIP 21547 & 2016-01-15 & DB\_H23 & 16x256 & 41.8 & 1.91 & 1.08 & 1.3 & 096.C-0241(G) \\
        HIP 21547 & 2016-12-11 & DB\_H23 & 64x54 & 25.3 & 1.97 & 1.12 & 1.5 & 198.C-0209(C)  \\
        HIP 21547 & 2016-12-12 & DB\_H23 & 64x72 & 45.0 & 0.84 & 1.09 & 5.7 & 198.C-0209(C) \\       
        \hline 
        \hline        
        \end{tabular}    
    \caption{51 Eri observation logs. \textbf{Notes:} $^a$: DIT corresponds to the detector integration time per frame, $\Delta$PA is the amplitude of the parallactic rotation, $\tau_0$ corresponds to the coherence time. $^b$: Values extracted from the updated DIMM info and averaged over the sequence. }
    \end{center}
    \label{tab:51eri_log}
\end{table*}

\begin{table*}
\centering
\begin{adjustbox}{width=\textwidth,center}
\begin{tabular}{|c|c|c|c||c|c|c|c||c|c|c|c||c|c|c|c|}			
\hline																			
\multicolumn{4}{|c||}{25/12/2015	} &	 \multicolumn{4}{|c||}{15/01/2016	}  & \multicolumn{4}{|c||}{11/12/2016	}  &	 \multicolumn{4}{|c|}{12/12/2016	    }\\
	\hline																					
SNR H2	& SNR ASDI & Found & \#C	& 	SNR	H2 & SNR ASDI & Found & \#C 	&SNR H2	& SNR ASDI & Found &\#C 	& SNR H2 & SNR ASDI & Found & \#C \\ \hline 
4.71& 4.3 & yes & 3  & 5.28 & 6.6 & yes & 3 & < 2.5 & < 4 & no & 4 & 2.69 & 4.1 & yes & 2 \\ 
\hline
\end{tabular}	
\end{adjustbox}
\caption{Results obtained on 51 Eridani.\label{tab:eridani}}
\end{table*}	

Our approach allows to find 51 Eri b in 3 out of the 4 datasets, and with an SNR lower than 5 (in the H2 band) in two cases. Moreover, a very small number of additional candidates are proposed (2 or 3). Unfortunately, none of the additional signals found by the classifier seems gravitationally bound to the star (i.e. detected at multiple epoch with a motion compatible with a bound object). We classify them as false positives.


\section{Conclusions }
Statistical approaches have proved very efficient to avoid self-subtraction when searching for exoplanets in ADI high contrast images, and to provide means to quantify the confidence of a detection. The SNR maps produced still contain too many artifacts related to background noise to simply identify planets with a threshold of SNR typically lower than $5\sigma$. But planetary signals and noise also leave specific patterns and shapes within the SNR map that can help discriminating them.
We have proposed a methodology using simple algorithmic techniques (edge-detection, regression and clustering) to help separating noise and planetary signals in these SNR maps. We demonstrated that the proposed methodology can considerably reduce the number of false positives and even improve detection in some cases (see for instance the case study of 51 Eridani). Moreover, it is well suited to learning with small data-sets (limited number of samples for the learning compared to current need of deep learning techniques) since it relies on dedicated and informative features of the application domain. This also helps explaining the results because the features have a meaning for the user. We now mostly intend to generalize it to spectroscopic data and test it on a larger scale. 

\label{sec:algorithm_detection}
\section*{Acknowledgements}


This project is supported in part by the European Research Council (ERC) under the European Union's Horizon 2020 research and innovation programme (COBREX; grant agreement n° 885593) as well as the CNRS (Mission pour les Initiatives Transverses et Interdisciplinaire: MITI).

\section{Data Availability Statement}

The data as well as detailed results are available  on a web page \footnote{\url{https://pagesperso.g-scop.grenoble-inp.fr/~catussen/exoplanet/report/}} and upon request.

\bibliographystyle{mnras}
\bibliography{bibliography} 


\appendix
\section{Features}

In the following we denote the maximum, mean and standard deviation over a set of reals numbers $X \subset \mathbb{R}$ by: $\max(X)$, $\mu(X)$ and $\sigma(X)$.

A stamp $S$ is a small sub-image restricted to a $D\times D$ area. This sub-image gives two $D\times D$ matrices of pixels or SNR values extracted from the original image in the two wavelengths $H2$ and $H3$. In practice, we experimented with stamps of size $29\times29$ and $19\times19$ ($D = 29$ and $D=19$). For sake of simplicity, we assume $D$ is always odd so that the center of the stamp is a pixel (and not in between two pixels). To make it easier to generalize $D$ to other contexts, we recall that we were working with 12,270 milli arc-seconds per pixel and two wavelengths at $\lambda_1=1.593 \mu m$ and $\lambda_2=1.667 \mu m$. Consider pixel $u = (i,j)$ in the stamp $S$ ($u \in S$), we denote by:
\begin{itemize}
\item $s^{H2}_{u}$ (resp. $s^{H3}_{u}$): the SNR value of pixel $u$ in wavelength $H2$ (resp $H3$).
\item $g_{u}$: the norm of the SNR gradient at pixel $u$ in wavelength  $H2$.
\item $c$: the pixel at the center of the stamp \emph{i.e} $c = (\lfloor D/2 \rfloor, \lfloor D/2 \rfloor)$ .
\item $\mathcal{C}$: the set of nine pixels in a $3\times 3$ area at the center of the stamp \emph{i.e} $\mathcal{C} = \{(i,j) | i \in \llbracket \lfloor D/2 \rfloor-1,\lfloor D/2 \rfloor+1\rrbracket, j \in \llbracket \lfloor D/2 \rfloor-1, \lfloor D/2 \rfloor+1\rrbracket\}$
\end{itemize}
\begin{figure}[t]
  \includegraphics[width=\linewidth]{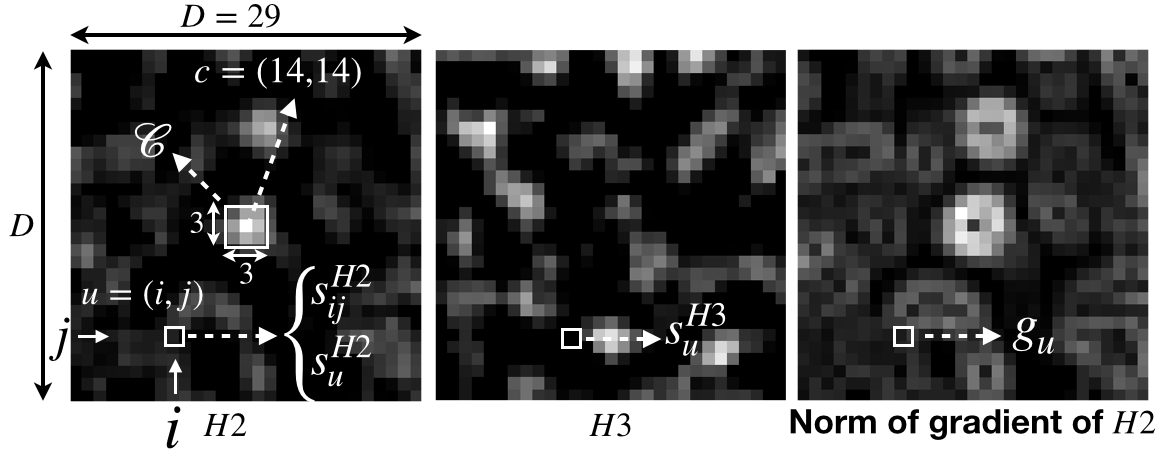}
  \caption{Summary of the notations for a stamp.}
  \label{fig:notationstamp}
\end{figure}

Note that we refer to pixel $u=(i,j)$ using its $i$ and $j$ coordinates when needed so that $s^{H2}_{u}$ can also be written $s^{H2}_{ij}$. Figure \ref{fig:notationstamp} gives a summary of the notations.

The features used are the following:
\begin{enumerate}
\item (\textbf{MeanSnr}). Mean SNR value in $H2$ and $H3$:
$$f^{H2}_1 = \mu (\{s^{H2}_u | u \in S\}), \qquad f^{H3}_1 = \mu (\{s^{H3}_u | u \in S\})$$
\item (\textbf{MaxCenteredSnr}). Max SNR value in $H2$ and $H3$ in the center $\mathcal{C}$:
$$f^{H2}_2 = \max (\{s^{H2}_u | u \in \mathcal{C} \}), \qquad f^{H3}_2 = \max (\{s^{H3}_u | u \in \mathcal{C} \})$$
\item  (\textbf{MaxGra}, \textbf{MeanGra},   \textbf{StdevGra}). Maximum, Mean and Standart deviation of the gradient in $H2$:
$$f_3 = \max(\{g_u | u \in S\}) $$
$$f_4 = \mu(\{g_u | u \in S\}) $$
$$f_5 = \sigma(\{g_u | u \in S\}) $$
\item (\textbf{MaxMin}). Consider the image defined for each pixel $u$ by the minimum SNR value between both wavelength: $\min(s^{H2}_u,s^{H3}_u)$. This \emph{minimum} image only keep the SNR quantities present in both wavelength. The idea is that a speckle systematically move between H2 and H3 whereas real companions \textbf{can be} present in both H2 and H3. We use the following feature:
$$f_6 = \max(\{\min(s^{H2}_u,s^{H3}_u) | u \in S\}) $$
\item  (\textbf{AiryFig}). To capture the presence of an Airy figure, we rely on an azimutal mean of SNR values. More precisely, let's denote $h_u(d)$ the mean SNR values in a ring of one pixel's width located at distance $d$ of pixel $u$. Figure \ref{fig:hd} shows $h_c(d)$ \emph{i.e} from the center $c$ of our example stamp. 
\begin{figure}
\centering
  \includegraphics[width=8cm]{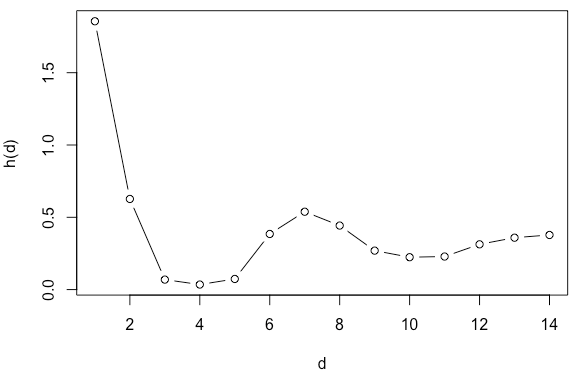}
  \caption{$h_c(d)$ function for the stamp presented figure \ref{fig:notationstamp}. The first dark ring can be seen around distance $4$ whereas the first white ring is around $7$. }
  \label{fig:hd}
\end{figure}

An Airy figure at the center $c$ of the stamp is expected to have a higher standard deviation $\sigma(c) = \sigma(\{h_{c}(d)| d \in \llbracket
1, D/2 \llbracket \})$ than random noise. This however assumes that the stamp is precisely centered. We therefore relax this constraint and take the best value among the nine central pixels. We use the following feature:
$$f_7 = \max(\{\sigma(u) | u \in \mathcal{C}\}) $$
Feature $f_6$ is computed for each wavelength which gives, in practice, two distinct features $f^{H2}_6$ and $f^{H3}_6$. \\
\item (\textbf{MeanSpec}). A speckle moves radially away from the center (the star) of the image depending on its distance to the center and the ratio $r=\frac{\lambda_{H3}}{\lambda_{H2}}= \frac{1.667}{1.593}$. Assuming the star is the origin of the coordinate system, if pixel $u=(i,j)$ in $H_2$ is part of a speckle and its center is located at coordinates $(x_i,y_j)$ in the star system, the point $ru = (rx_i,ry_j)$ in $H3$ is expected to have the same SNR intensity. Since $(rx_i,ry_j)$ does not necessarily match exactly the center of another pixel, we compute a weighted average of its four neighbors. Let $N(u)$ be the nine closest pixels of the real point $ru = (rx_i,ry_j)$, we compute $w(u)$ (where $d'(u,v)= \frac{1}{d(u,v)}$ is the inverse of the distance between pixel $u$ and $v$): 
$$w(u) = \frac{\sum_{v\in N(u)} d'(ru,v) \times s^{H3}_{v}}{\sum_{v \in N(u)} d'(ru,v)}$$

\begin{figure}
\centering
  \includegraphics[width=5cm]{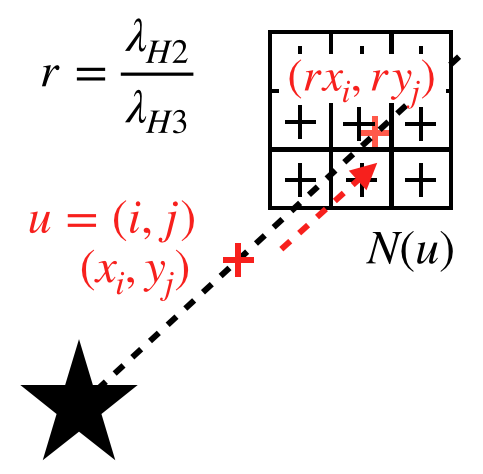}
  \caption{Motion of a speckle from H2 to H3}
  \label{fig:speckle}
\end{figure}

The following quantity $sp(u)$ is used to quantify how much the SNR intensity $s^{H2}_u$ is found at $(rx_i,ry_j)$ in $H3$ i.e is close to $w(u)$:
$$sp(u) = \frac{|s^{H2}_{u} - w(u)|}{\max(s^{H2}_{u},w(u))}$$
Finally the feature considers the 9 central pixels to identify a speckle:
$$f_8 = \mu(\{sp(u) | u \in \mathcal{C}\})$$
The previous caracterization of a speckle is only valid at minimum and maximum distance from the star. We therefore introduce an indicative feature:
$$f_9 = \left\{ \begin{array}{ll}
1 & \textrm{ if }d_{min} \leq d(c,star) \leq d_{max}\\
0 & \textrm{ otherwise }
\end{array}\right. $$
We used the values (in arc-seconds) of $d_{min} = 0.2"$ and $d_{max} = 1.7"$.\\

\item (\textbf{MeanNoise, StdevNoise}). At greater distances from the star ($> d_{max}$) a companion object can sometimes disappear from $H2$ to $H3$. The quantity $n(u)$ measures how much of the intensity in $H2$ at pixel $u$ disappears in $H3$:
$$n(u) = max(0,\frac{s^{H2}_{u} -s^{H3}_{u}}{\max(s^{H2}_{u},s^{H3}_{u})})$$
We use the mean and the standart deviation of this quantity, at the center of the stamp, as features:
$$f_{10} = \mu(\{n(u) |u \in \mathcal{C}\}), \qquad f_{11} = \sigma(\{n(u) |u \in \mathcal{C}\})$$

\item (\textbf{Dist}). The distance of the center of the stamp to the star is also included in the features and denoted $f_{12}$.

\end{enumerate}

\section{Annexe}

\begin{figure*}
\begin{subfigure}
  \centering
  \includegraphics[width=.49\linewidth]{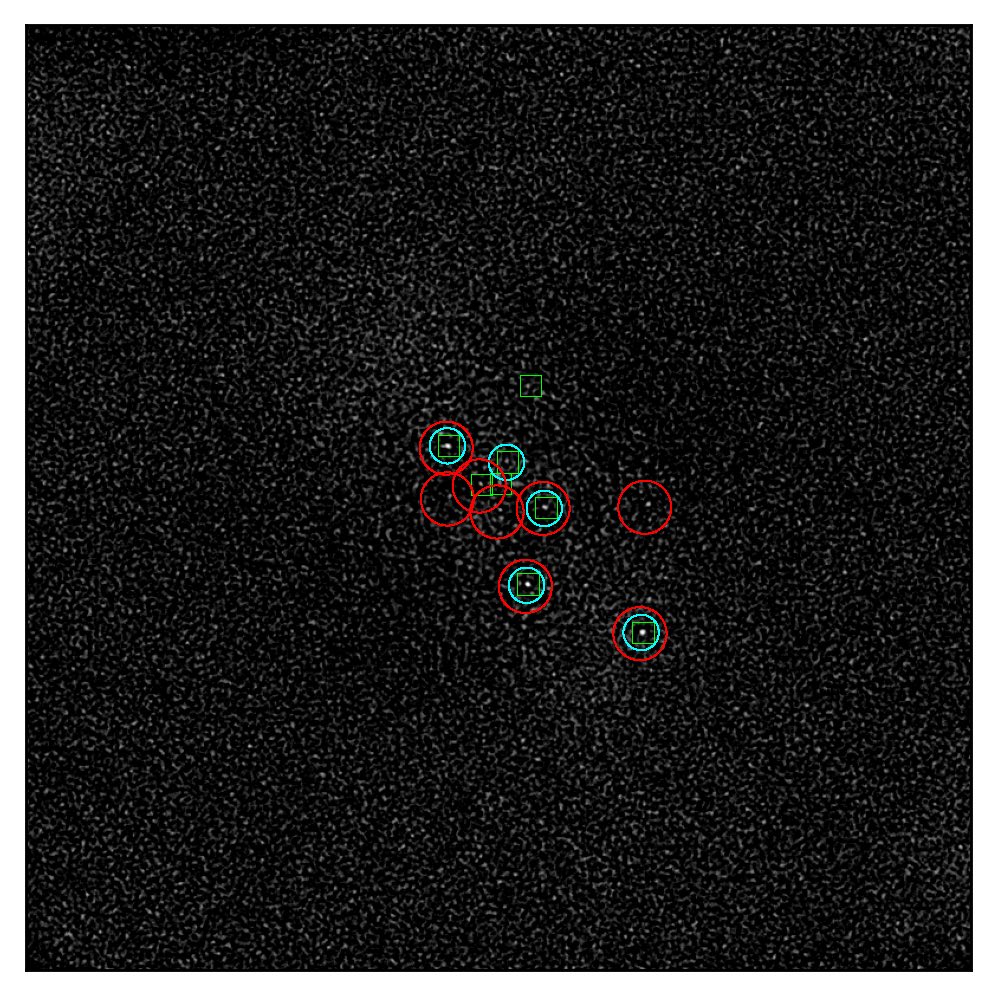}
\end{subfigure}%
\begin{subfigure}
  \centering
  \includegraphics[width=.49\linewidth]{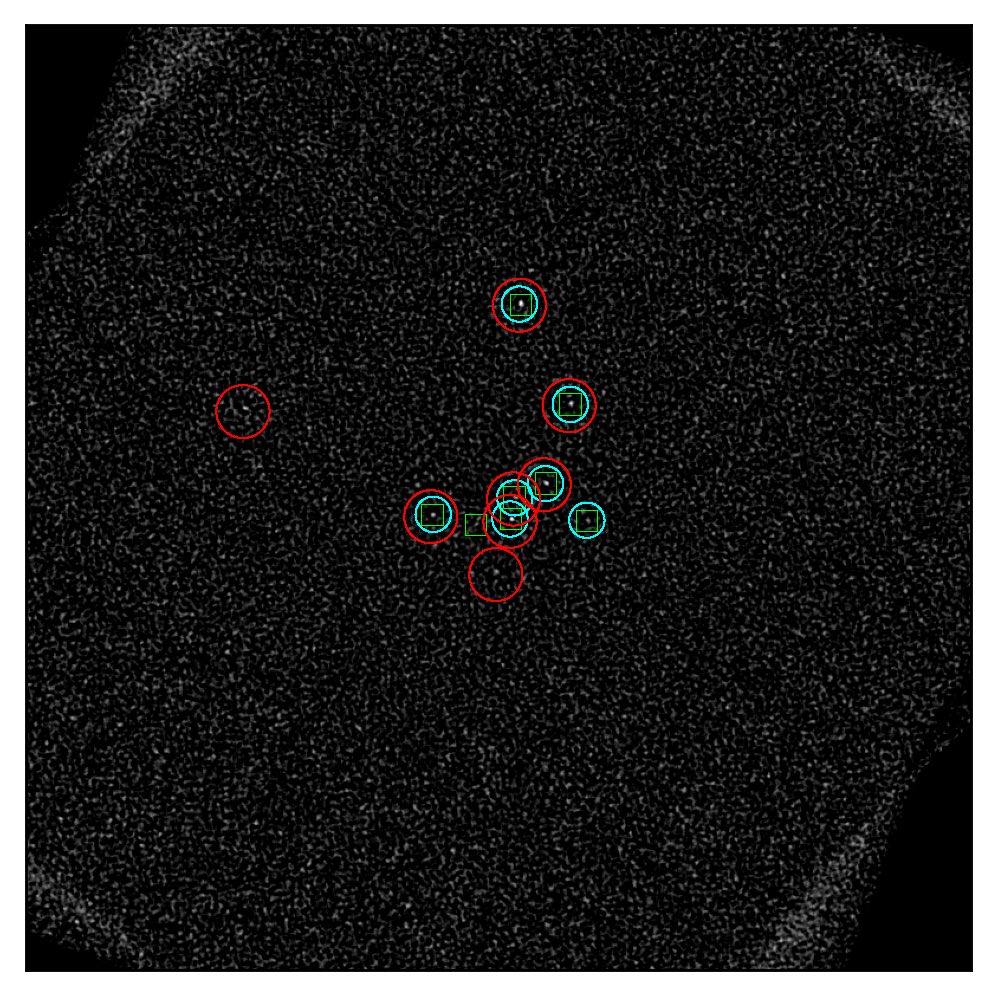}
\end{subfigure}
\begin{subfigure}
  \centering
  \includegraphics[width=.49\linewidth]{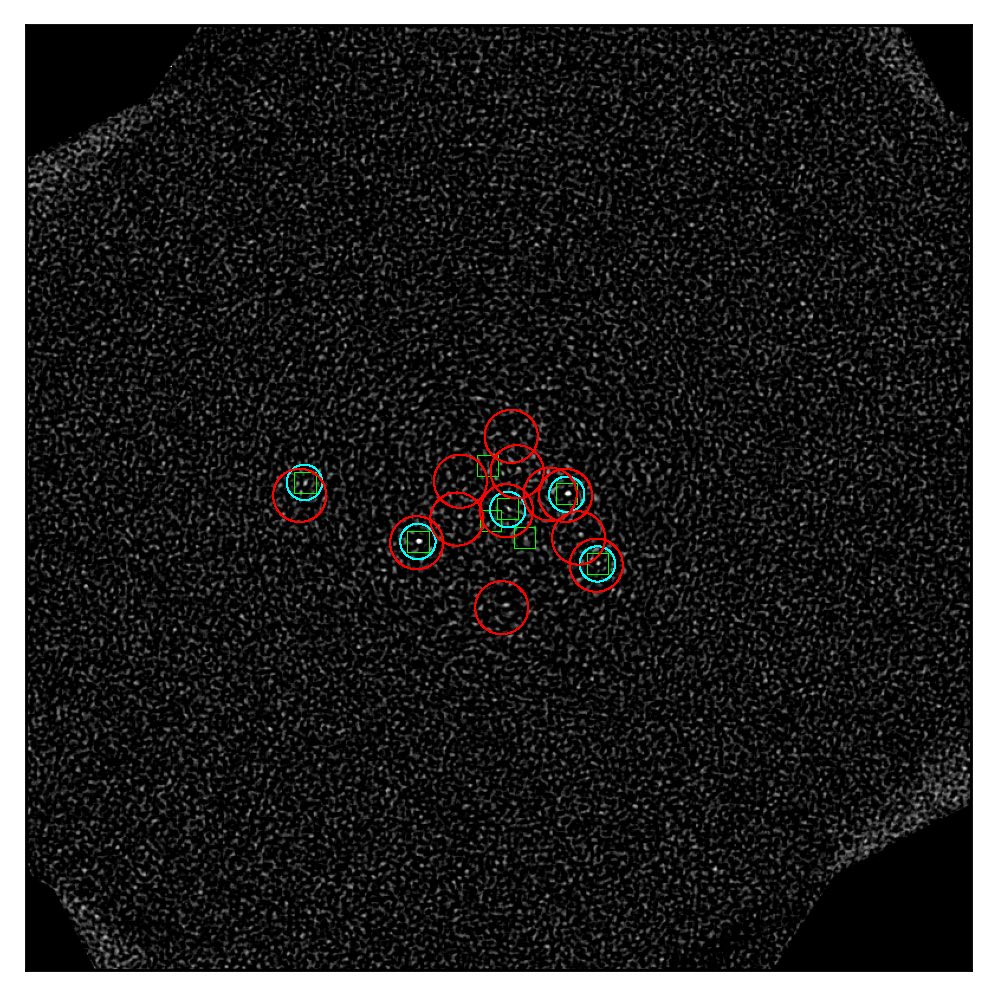}
\end{subfigure}
\begin{subfigure}
  \centering
  \includegraphics[width=.49\linewidth]{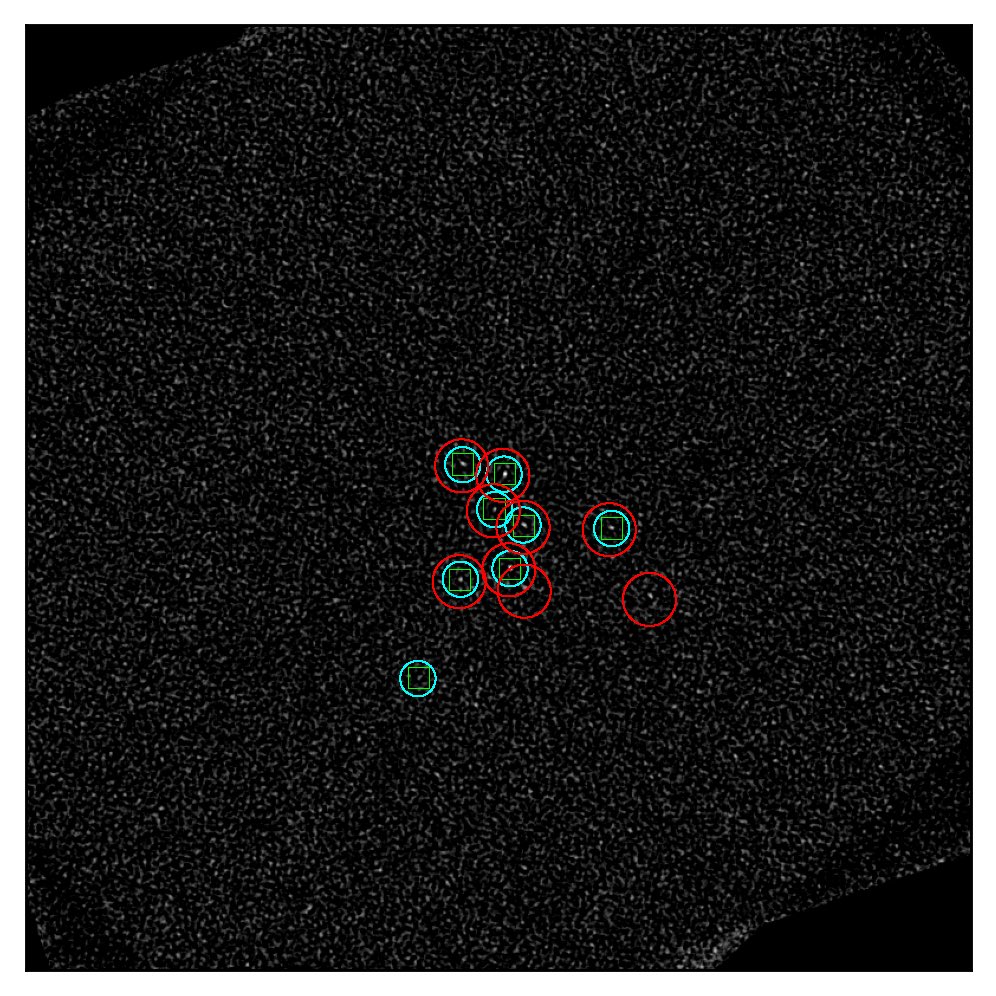}
\end{subfigure}
\caption{Visualization of the results on BT1. An injected exoplanet is diplayed by a green square, a candidate proposed by PACO ASDI 5 is a little blue circle, a candidate proposed by regL is a large red circle. Top left: HD108767B, Top right: HIP1993, Bottom left: HIP12394, Bottom right: HIP107345.}
\label{fig:bt1}
\end{figure*}

\begin{figure*}
\begin{subfigure}
  \centering
  \includegraphics[width=.49\linewidth]{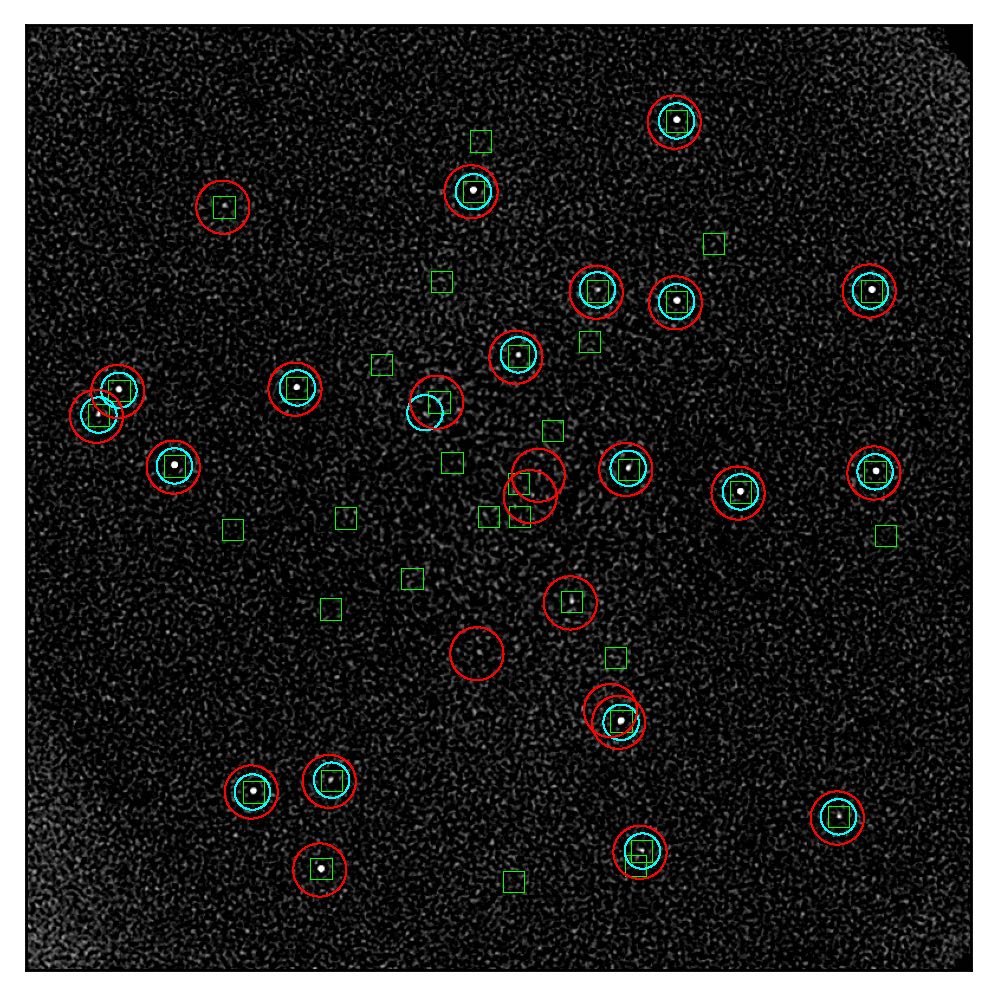}
\end{subfigure}%
\begin{subfigure}
  \centering
  \includegraphics[width=.49\linewidth]{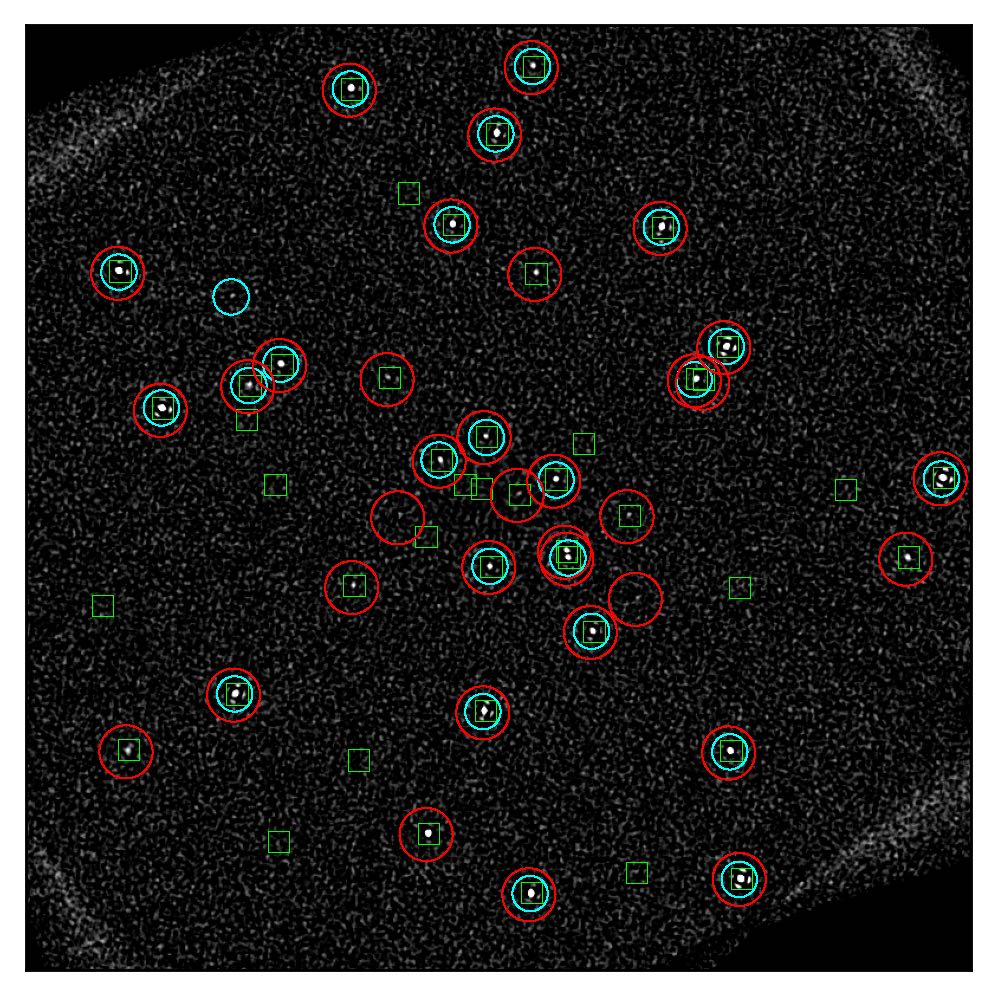}
\end{subfigure}
\begin{subfigure}
  \centering
  \includegraphics[width=.49\linewidth]{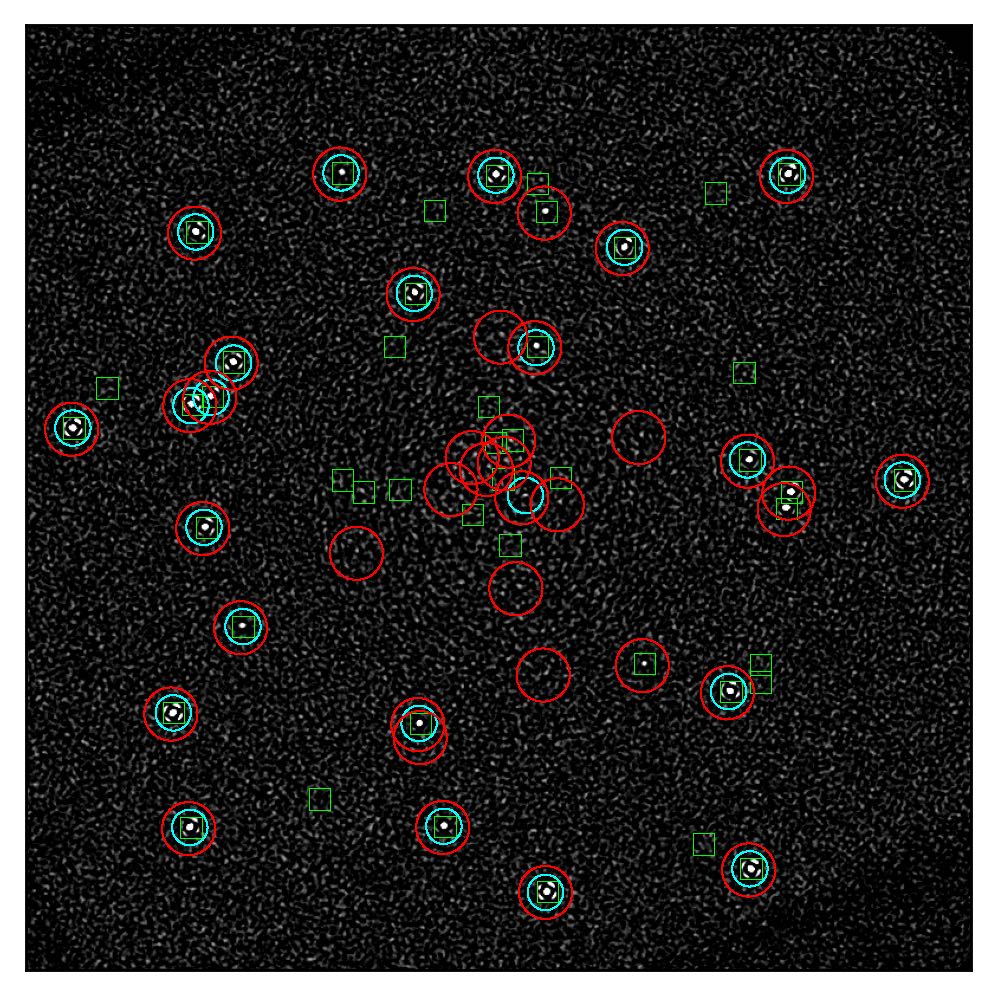}
\end{subfigure}
\begin{subfigure}
  \centering
  \includegraphics[width=.49\linewidth]{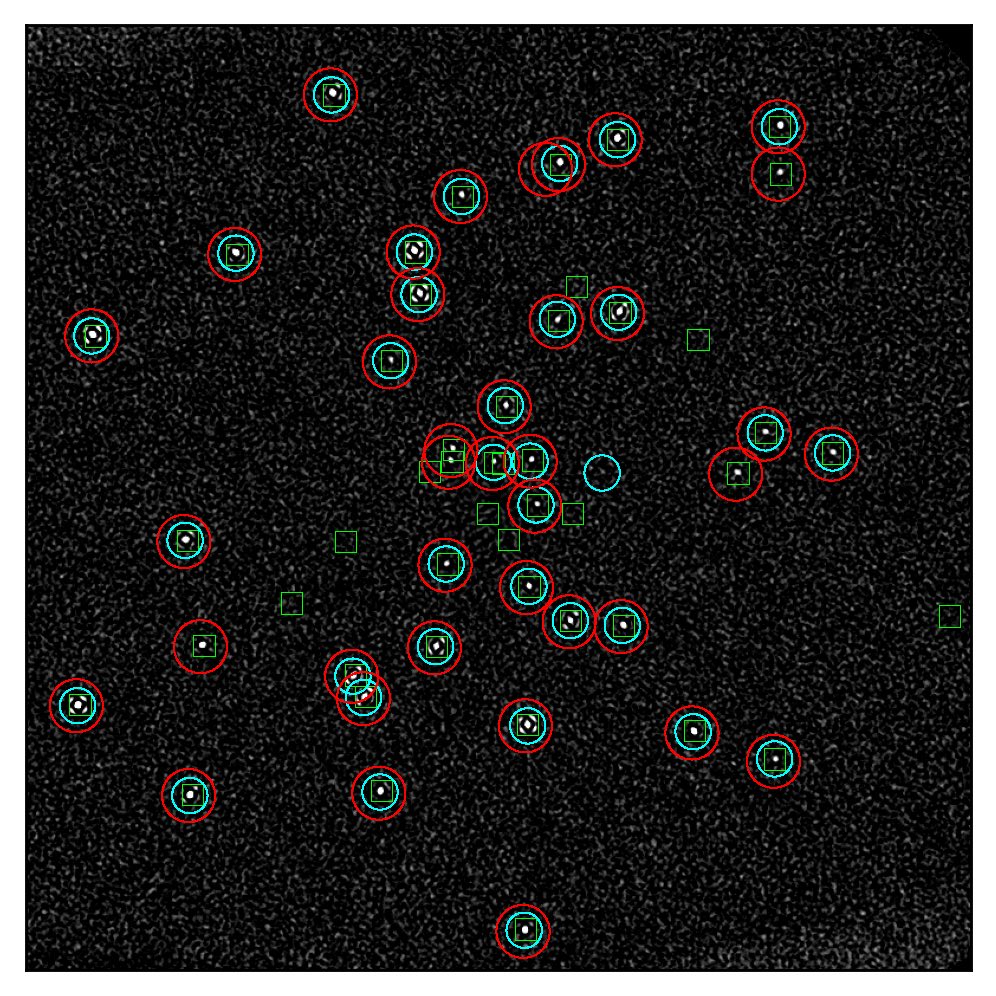}
\end{subfigure}
\caption{Visualization of the results on \textbf{BT2}. An injected exoplanet is diplayed by a green square, a candidate proposed by PACO ASDI 5 is a little blue circle, a candidate proposed by regL is a large red circle. Top left: HD108767B, Top right: HIP1993, Bottom left: HIP12394, Bottom right: HIP107345.}
\label{fig:bt2}
\end{figure*}

\begin{figure*}
\begin{subfigure}
  \centering
  \includegraphics[width=.49\linewidth]{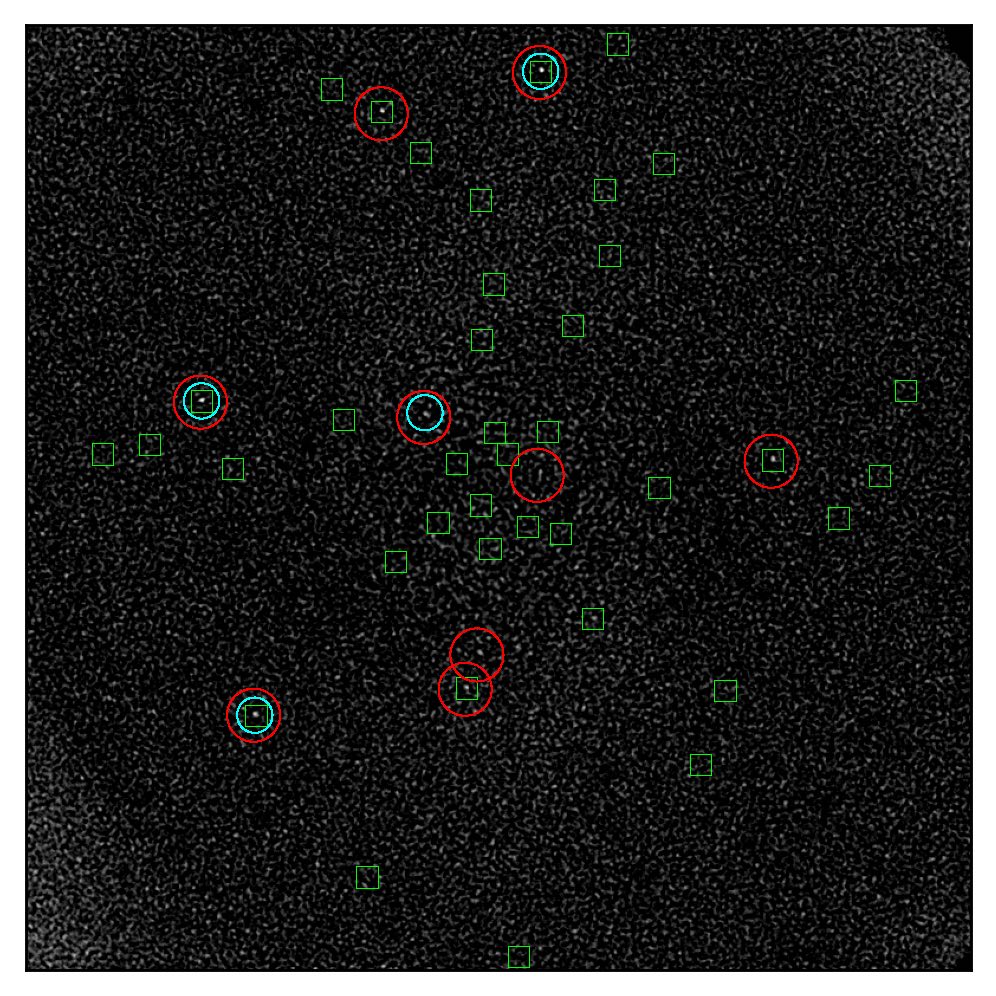}
\end{subfigure}%
\begin{subfigure}
  \centering
  \includegraphics[width=.49\linewidth]{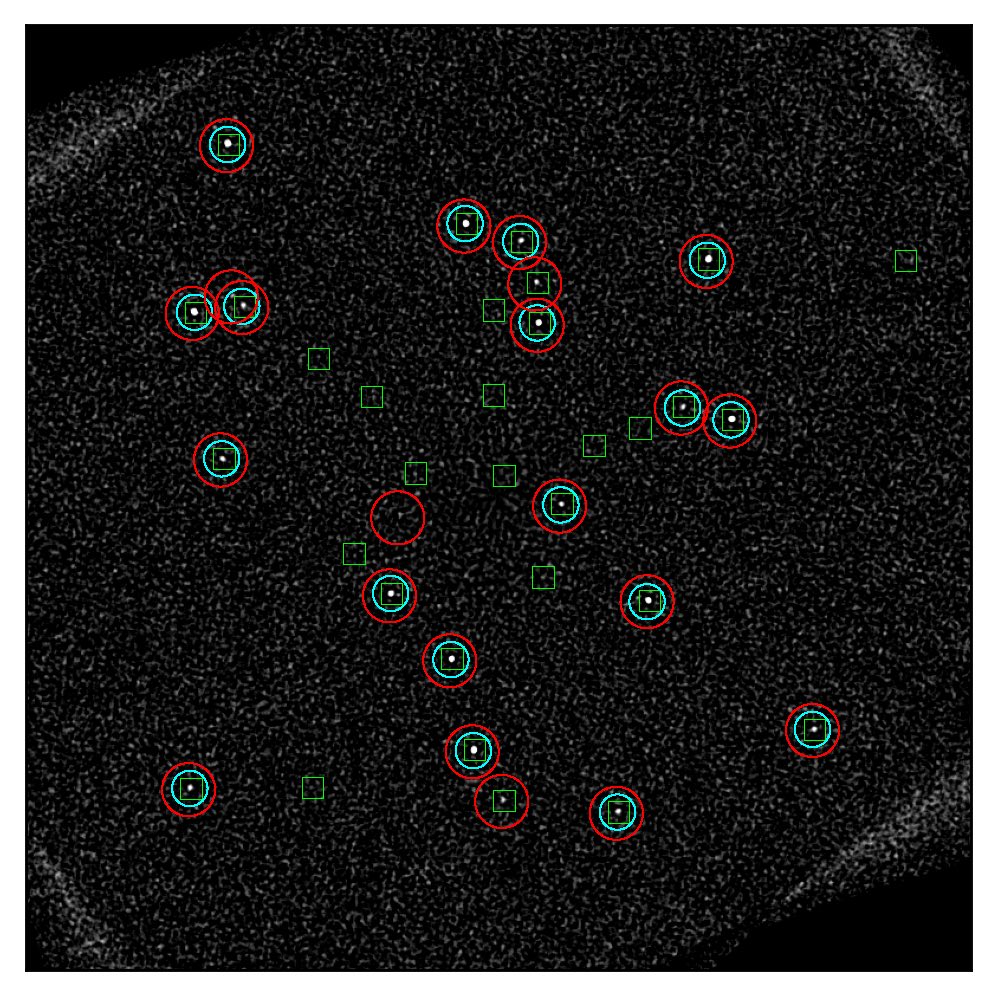}
\end{subfigure}
\begin{subfigure}
  \centering
  \includegraphics[width=.49\linewidth]{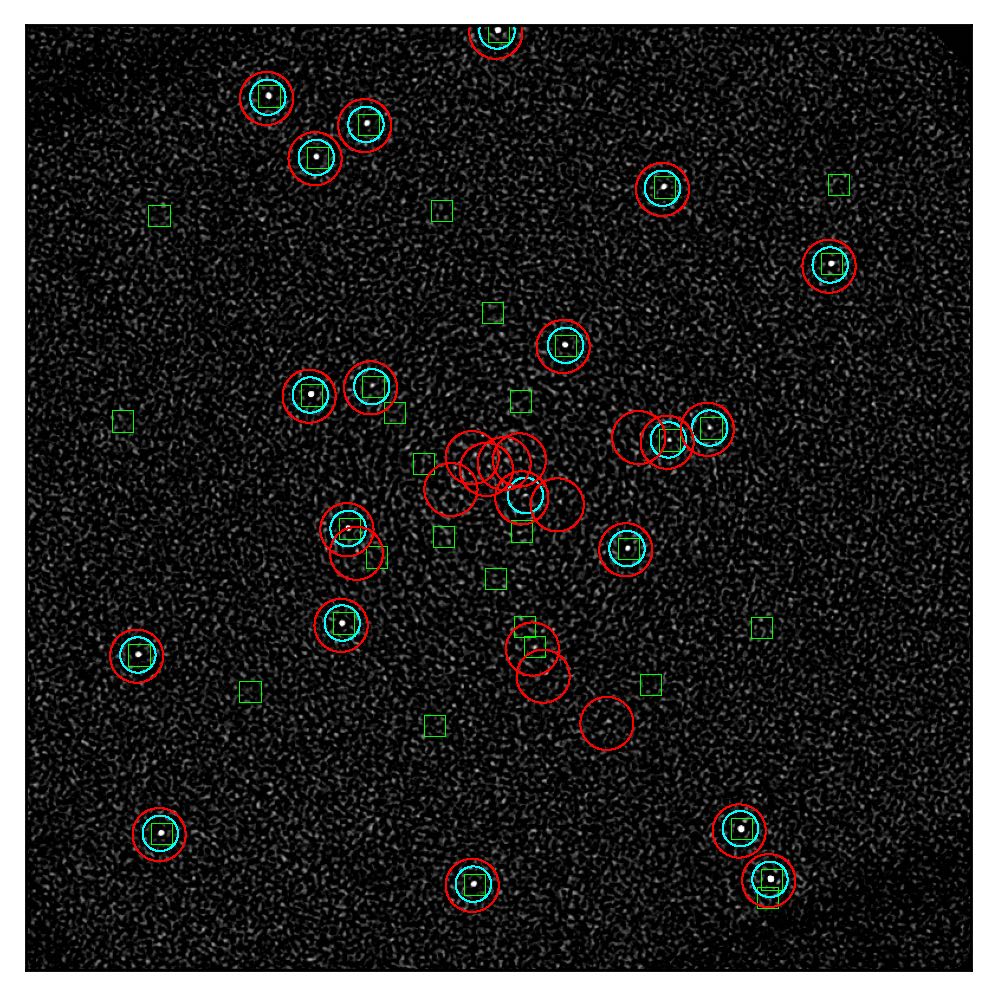}
\end{subfigure}
\begin{subfigure}
  \centering
  \includegraphics[width=.49\linewidth]{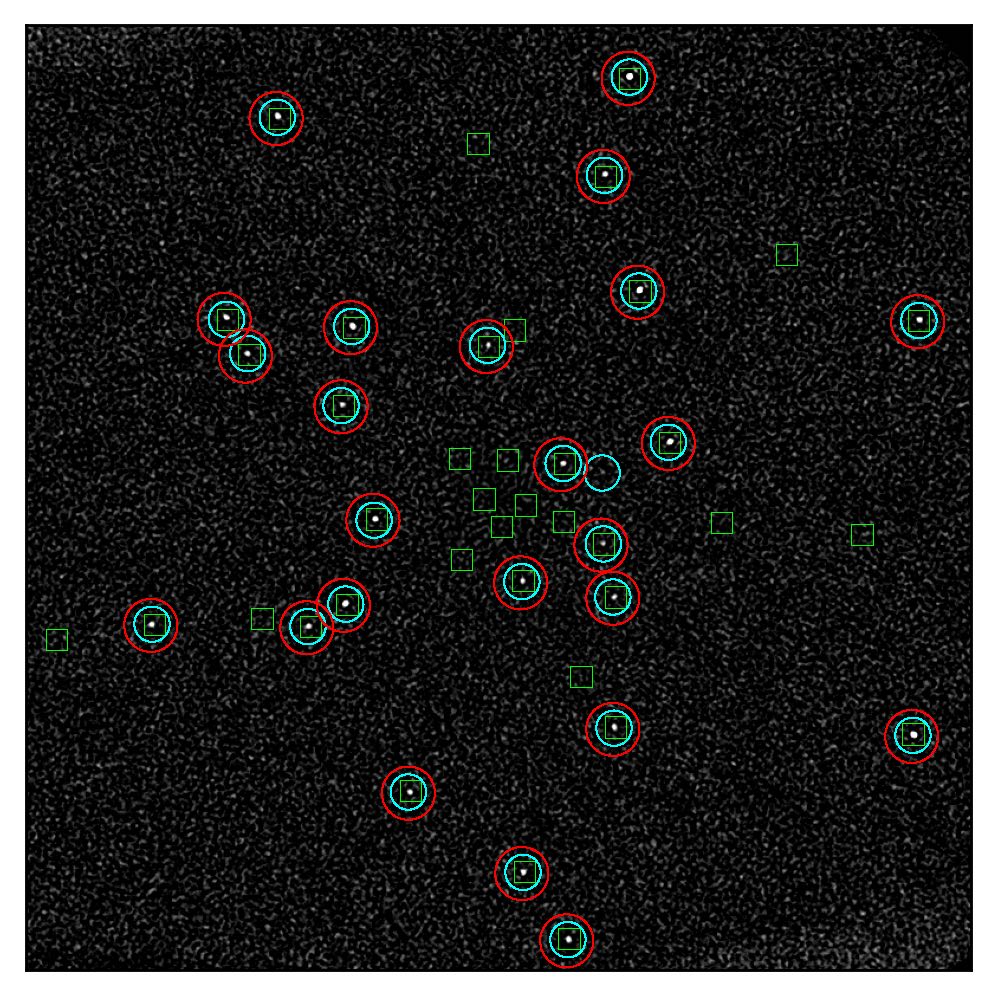}
\end{subfigure}
\caption{Visualization of the results on \textbf{BT3}. An injected exoplanet is diplayed by a green square, a candidate proposed by PACO ASDI 5 is a little blue circle, a candidate proposed by regL is a large red circle. Top left: HD108767B, Top right: HIP1993, Bottom left: HIP12394, Bottom right: HIP107345.}
\label{fig:bt3}

\end{figure*}

Figures \ref{fig:bt1}, \ref{fig:bt2}, \ref{fig:bt3} present the results directly on the SNR maps. The exoplanets injected are located as little green squares, the results (candidates proposed) of PACO ASDI 5 are shown as little blue circles whereas the results of our approach regL are displayed as large red circles. For instance, exoplanets found only by PACO ASDI 5 therefore appear as a little square within a little circle. Reversely, a square within a large circle is a target identified by regL. Some targets are not found by any of the methods.

\end{document}